\documentclass[12pt]{article}%
\usepackage[nosort]{cite}
\usepackage[usenames, dvipsnames]{xcolor}
\usepackage{graphicx}
\usepackage{multicol}
\usepackage{amsfonts}
\usepackage{amssymb}
\usepackage{amsmath}
\usepackage{heck}
\usepackage{afterpage}
\usepackage{setspace}
\usepackage{verbatim}
\usepackage{longtable}
\usepackage{float}
\usepackage{subcaption}
\usepackage{epsfig}
\usepackage{enumerate}
\usepackage{epstopdf}
\usepackage[enableskew, vcentermath]{youngtab}
\usepackage{adjustbox}
\usepackage{multirow}
\usepackage{tikz}
\usepackage[margin=1in]{geometry}
\usepackage{titletoc}
\usepackage{hyperref}
\usepackage{gensymb}
\usepackage{mathtools}%
\usepackage{tikz-cd}
\providecommand{\U}[1]{\protect\rule{.1in}{.1in}}
\pdfoutput=1
\newsavebox{\mysavebox}

\hypersetup{colorlinks,citecolor=black,filecolor=black,linkcolor=black,urlcolor=black}
\usetikzlibrary{decorations.markings}

\numberwithin{equation}{section}

\usetikzlibrary{chains}
\allowdisplaybreaks
\tikzset{node distance=2em, ch/.style={circle,draw,on chain,inner sep=2pt},chj/.style={ch,join},every path/.style={shorten >=4pt,shorten <=4pt},line width=1pt,baseline=-1ex}

\newcommand{\Ni}{N_{\rm inf}}
\newcommand{\dNi}{\dot{N}_{\rm inf}}
\newcommand{\ddNi}{\ddot{N}_{\rm inf}}

\newcommand{\Nd}{N_{\rm DE}}
\newcommand{\dNd}{\dot{N}_{\rm DE}}

\newcommand{\ba}{\begin{eqnarray}}
\newcommand{\ea}{\end{eqnarray}}

\newcommand{\be}{\begin{equation}}
\newcommand{\ee}{\end{equation}}

\newcommand{\dd}{\, {\rm d}}

\newcommand{\ve}{\varepsilon}

\newcommand{\ls}{\ell_{\rm s}}

\tikzstyle{startstop} = [rectangle, rounded corners, minimum width=3cm, minimum height=1cm,text centered, draw=black, fill=blue!10]
\tikzstyle{startstop} = [rectangle, rounded corners, minimum width=3cm, minimum height=1cm,text centered, draw=black, fill=blue!10]
\tikzstyle{io} = [trapezium, trapezium left angle=70, trapezium right angle=110, minimum width=3cm, minimum height=1cm, text centered, draw=black, fill=blue!30]
\tikzstyle{process} = [rectangle, minimum width=3cm, minimum height=1cm, text centered, draw=black, fill=orange!30]
\tikzstyle{decision} = [diamond, minimum width=3cm, minimum height=1cm, text centered, draw=black, fill=green!30]
\tikzstyle{arrow} = [thick,->,>=stealth]
\tikzset{->-/.style={decoration={
  markings,
  mark=at position #1 with {\arrow[scale=2.4]{>}}},postaction={decorate}}}
\makeatletter \@addtoreset{equation}{section} \makeatother

\begin{document}

\date{January 2019}

\title{Pixelated Dark Energy}

\institution{PENN}{\centerline{Department of Physics and Astronomy, University of Pennsylvania, Philadelphia, PA 19104, USA}}

\authors{Jonathan J. Heckman\footnote{e-mail: {\tt jheckman@sas.upenn.edu}},
Craig Lawrie\footnote{e-mail: {\tt craig.lawrie1729@gmail.com}},\\[4mm]
Ling Lin\footnote{e-mail: {\tt lling@physics.upenn.edu}}, Jeremy Sakstein\footnote{e-mail: {\tt sakstein@physics.upenn.edu}},
and Gianluca Zoccarato\footnote{e-mail: {\tt gzoc@sas.upenn.edu}}}

\abstract{\noindent We study the phenomenology of a recent string construction
with a quantum mechanically stable dark energy. A mild supersymmetry
protects the vacuum energy but also allows $O(10 - 100)$ TeV scale
superpartner masses. The construction is holographic in the sense that the 4D spacetime
is generated from ``pixels'' originating from five-branes wrapped over metastable five-cycles
of the compactification. The cosmological constant scales as $\Lambda \sim 1/N$ in the pixel number.
An instability in the construction leads to cosmic expansion. This also causes
more five-branes to wind up in the geometry, leading to a slowly decreasing
cosmological constant which we interpret as an epoch of inflation followed by \mbox{(pre-)}heating
when a rare event occurs in which the number of pixels increases by an order one fraction.
The sudden appearance of radiation triggers an exponential increase in the number of pixels.
Dark energy has a time varying equation of state with $w_a=-3\Omega_{m,0}(1+w_0)/2$,
which is compatible with current bounds, and could be constrained
further by future data releases. The pixelated nature of the Universe
also implies a large-$l$ cutoff on the angular power spectrum of
cosmological observables with $l_{\rm max} \sim O(N)$.
We also use this pixel description to study the thermodynamics of de Sitter space, finding
rough agreement with effective field theory considerations.}

\maketitle

\setcounter{tocdepth}{2}

\tableofcontents


\newpage

\section{Introduction} \label{sec:INTRO}

Cosmology is a promising arena for probing quantum gravity.  Developing robust
UV-complete models for dark energy as well as the origin of the observed power
spectrum of primordial density perturbations provides a potential window into
the inner structure of string theory, the only known consistent theory of
quantum gravity.

Even so, bottom up considerations suggest a seemingly endless variety of low
energy effective field theories which can consistently match observations. For
example, there are a plethora of single field slow-roll inflation models (and
multi-field generalizations thereof) and several dark energy candidates such
as quintessence \cite{Copeland:2006wr}, K-essence, and infrared modifications
of gravity
\cite{Clifton:2011jh,Joyce:2014kja,Koyama:2015vza,Burrage:2016bwy,Burrage:2017qrf,Sakstein:2018fwz}.
One might hope that UV motivated models generated from string theory would
preferentially favor some such scenarios. There are, however, a large number
of de Sitter-like constructions (see e.g.~\cite{Maloney:2002rr, Kachru:2003aw,
Balasubramanian:2005zx, Westphal:2006tn, Dong:2010pm, Rummel:2011cd,
Blaback:2013fca, Cicoli:2013cha, Cicoli:2015ylx}) as well as an apparent
landscape of low energy effective field theories.\footnote{ Even if this turns
  out to be the case, it is still important to figure out \textit{which} of
  the many (but still \textit{finite}) string constructions actually describes
the observed Universe.} This would seem to suggest that cosmology may not
actually be a particularly good way to test string theory.

On the other hand, it has also been appreciated for some time that such
constructions involve a large number of moving parts, which are not always
completely under perturbative control.  It is common to appeal to the matching
of effective field theories (see e.g.~\cite{Polchinski:2015bea,
Kachru:2018aqn}), and the expectation is that suitable UV boundary conditions
are compatible with a given string construction.  There are also well-known
critiques of de Sitter constructions, e.g.~references \cite{Maldacena:2000mw,
  Townsend:2003qv, Hertzberg:2007wc, Covi:2008ea, Caviezel:2008tf,
Caviezel:2009tu, deCarlos:2009fq, Wrase:2010ew, Shiu:2011zt, Green:2011cn,
Gautason:2012tb, Bena:2014jaa, Kutasov:2015eba, Quigley:2015jia,
Dasgupta:2014pma, Junghans:2016abx, Junghans:2016uvg, Andriot:2016xvq,
Moritz:2017xto, Sethi:2017phn, Andriot:2017jhf, Danielsson:2018ztv} and there
has recently been much discussion in the context of the Swampland conjecture
for the absence of de Sitter space \cite{Obied:2018sgi}. For some recent
examples of the literature see references
\cite{Agrawal:2018mkd,Agrawal:2018own,Andriot:2018wzk,Branchina:2018xdh,Dvali:2018fqu,Banerjee:2018qey,
  Aalsma:2018pll,Achucarro:2018vey,Garg:2018reu,Lehners:2018vgi,Kehagias:2018uem,Dias:2018ngv,Denef:2018etk,
  Colgain:2018wgk,Roupec:2018mbn,Andriot:2018ept,Ghosh:2018fbx,Matsui:2018bsy,Ben-Dayan:2018mhe,Chiang:2018jdg,
  Heisenberg:2018yae,Damian:2018tlf,Conlon:2018eyr,Kinney:2018nny,Dasgupta:2018rtp,Cicoli:2018kdo,
  Akrami:2018ylq,Nakai:2018hhf,Heisenberg:2018rdu,Murayama:2018lie,Marsh:2018kub,Brahma:2018hrd,Choi:2018rze,Das:2018hqy,
  Danielsson:2018qpa,Wang:2018duq,Brandenberger:2018wbg,DAmico:2018mnx,Han:2018yrk,Moritz:2018ani,Bena:2018fqc,
  Brandenberger:2018xnf,Dimopoulos:2018upl,Ellis:2018xdr,Lin:2018kjm,Hamaguchi:2018vtv,Kawasaki:2018daf,Motaharfar:2018zyb,
  Odintsov:2018zai,Ashoorioon:2018sqb,Antoniadis:2018ngr,Das:2018rpg,Ooguri:2018wrx,Wang:2018kly,Fukuda:2018haz,Buratti:2018onj,
  Hebecker:2018vxz,Gautason:2018gln,Olguin-Tejo:2018pfq,Garg:2018zdg,Dvali:2018jhn,Park:2018fuj,Blaback:2018hdo,Schimmrigk:2018gch,
Lin:2018rnx,Kim:2018mfv,Agrawal:2018rcg,Yi:2018dhl,Chiang:2018lqx,Grimm:2018cpv,Dvali:2018txx,Dvali:2018dce,Thompson:2018ifr,
Russo:2018akp,Cheong:2018udx,Elizalde:2018dvw,Ibe:2018ffn,Blanco-Pillado:2018xyn,Tosone:2018qei,Holman:2018inr,Junghans:2018gdb,
Emelin:2018igk,Klaewer:2018yxi,Banlaki:2018ayh,Andriot:2018mav,Acharya:2018deu,Herdeiro:2018hfp,Bonnefoy:2018mqb,Kinney:2018kew,
Hertzberg:2018suv,Cordova:2018dbb,Lin:2018edm,Buratti:2018xjt,Hamada:2018qef,Gonzalo:2018guu,Bastero-Gil:2018yen,
Corvilain:2018lgw,Scalisi:2018eaz,Seo:2018abc,Craig:2018yvw,Dvali:2018ytn,Abel:2018zyt,Raveri:2018ddi,Cai:2018ebs,
Arciniega:2018tnn,Kallosh:2019axr,Kiritsis:2019wyk,Weissenbacher:2019mef,Antoniadis:2019doc,Bjorkmo:2019aev,Weissenbacher:2019bfb}.

By a similar token, similar issues arise in the search for stringy models of
inflation, in part because inflation can be viewed as a nearly de Sitter-like
phase. Again, the number of moving parts in most stringy constructions makes
it challenging to extract robust predictions for a given class of models.

Much of the debate on the existence of de Sitter vacua in string theory has
centered on whether it is possible to find examples of scalar potentials with
a (small) positive energy density at a metastable minimum. These scalars
originate from the remnants of 10D quantum gravity in our 4D world, the
so-called moduli of the internal geometry.  Stabilizing these moduli---finding
models with a suitable potential---is a challenging question, both from a
practical perspective as well as from a conceptual standpoint.

From a practical standpoint, it is often necessary to consider
non-perturbative effects.  For such ingredients to play a prominent r\^{o}le
implies that perturbation theory in the original parameters is unavailable
(see e.g. \cite{Dine:1985he}).  In such cases alternative perturbative
expansions in a suitable effective field theory must instead be used. This can
be delicate to arrange.

From a conceptual standpoint, it is also not entirely clear whether it is even
appropriate to generalize the language of scalar potentials in Minkowski space to de
Sitter space \cite{Banks:2003vp}. The asymptotic causal structure of
de Sitter space and de Sitter-like spacetimes means one cannot really ``freeze
VEVs'' as one would do in Minkowski or Anti-de Sitter space, especially in the
long distance limit where effective field theory is supposed to apply.

The common thread in all of these issues is the focus on fundamental scalars,
and whether they can be suitably ``stabilized.'' If they can, then one should
expect pure de Sitter vacua in string theory, and a constant equation of state
for dark energy $w=P/\rho=-1$. Perturbations away from such scenarios would
also generate models of inflation.  If scalars cannot be stabilized at
positive energy density, one should not expect the equation of state for dark
energy to be exactly constant \cite{Agrawal:2018own}, and based on various
conjectures (with varying degrees of evidence), it has even been suggested
that slow roll inflation is incompatible with string theory
\cite{Ooguri:2018wrx}.

Of course, finding self-consistent models of time-dependent dark energy which
are not ruled out by other constraints is potentially even more problematic
than just finding de Sitter vacua. Quintessence models, where a scalar slowly
rolling down a potential drives a quasi-de Sitter expansion, are highly
fine-tuned from the viewpoint of the low energy EFT \cite{Hertzberg:2018suv}
(even more so than a cosmological constant). More natural models where the
scalar couples to all particle species are typically ruled out by fifth-force
searches \cite{Kapner:2006si,Burrage:2016bwy,Burrage:2017qrf} and solar system
tests of gravity \cite{Sakstein:2017pqi}. It can also be difficult to realize
the low energy scales required using stringy constructions
\cite{Cicoli:2018kdo}.

In this paper we propose a stringy alternative to such considerations. The
starting point is F-theory on a $\mathrm{Spin}(7)$ background;
$\mathrm{Spin}(7)$ manifolds have recently been studied in a string theory
context in
\cite{Bonetti:2013fma,Bonetti:2013nka,Braun:2018joh,Heckman:2018mxl}. As
proposed in reference \cite{Heckman:2018mxl}, the resulting 4D spacetime has
topology $\mathbb{R}_{\text{time}} \times S^3$ (i.e. the Universe is closed)
and represents a cosmology in which dark energy (with equation of state $w =
-1$) balances against positive spatial curvature and a stiff fluid (with
equation of state $w = +1$). The stiff fluid arises from a background
three-form flux (magnetic flux associated with an axion) which is quantized in
integer units $N$. The various energy densities scale with the number of flux
units, $N$, as
\begin{equation}
\rho_{\text{curvature}} \sim -\frac{1}{a^2}, \quad \rho_{\Lambda} \sim
\frac{1}{N}, \quad \rho_{\text{stiff}} \sim \frac{N^2}{a^6} \,,
\end{equation}
with $a$ being the scale factor of a closed FRW spacetime:
\begin{equation}
  \dd s^{2}_{\rm FRW} = -\dd t^2 + a^2(t) \dd \Omega_{S^3}^2 \,.
\end{equation}
Compactifying F-theory on a $\mathrm{Spin}(7)$ background gives rise to a supersymmetric ground state, but all finite energy excitations have a non-supersymmetric spectrum \cite{Heckman:2018mxl}.  This leads to a
UV-complete model of dark energy in which zero point (vacuum) energies from
all quantum sectors automatically cancel.  In \cite{Heckman:2018mxl} this was
referred to as ``$\mathcal{N} = 1/2$ supersymmetry" and we continue to do so
in this work. This implies that the cosmological constant is both
UV-insensitive and radiatively stable, and therefore there is no cosmological
constant problem.\footnote{The reader concerned with phase transitions should
not be. These are concerns for infrared modifications of gravity that attempt
to \emph{degravitate} an arbitrarily large cosmological constant (e.g.
\cite{Dvali:2007kt,Khoury:2018vdv}) because such theories may not be able to
degravitate the vacuum energy both before and after a transition. In our
construction, the vacuum energy at early times is larger than today due to
symmetry restoration but this is never problematic since the phase transitions
occur while the large vacuum energy is still sub-dominant to radiation.} From
the low-energy perspective, the cosmological constant is a measurable quantity
just like any UV-insensitive and radiatively stable quantity. Given that we
also have an expression for it coming from UV stringy physics, this
measurement sets the number of flux units, or ``pixels'', $N$, today to be $\sim 10^{120}$. We
expect that since the $S^3$ is actually flat with respect to a twisted spin
connection that standard issues with Freund--Rubin compactification such as
having large extra dimensions are bypassed. Said differently, there appears to
be no issue with generating a small value for the 4D Newton's constant.

\begin{figure}[t!]
\begin{center}
\includegraphics[trim={0cm 2.5cm 0cm 3cm},clip,scale=0.5]{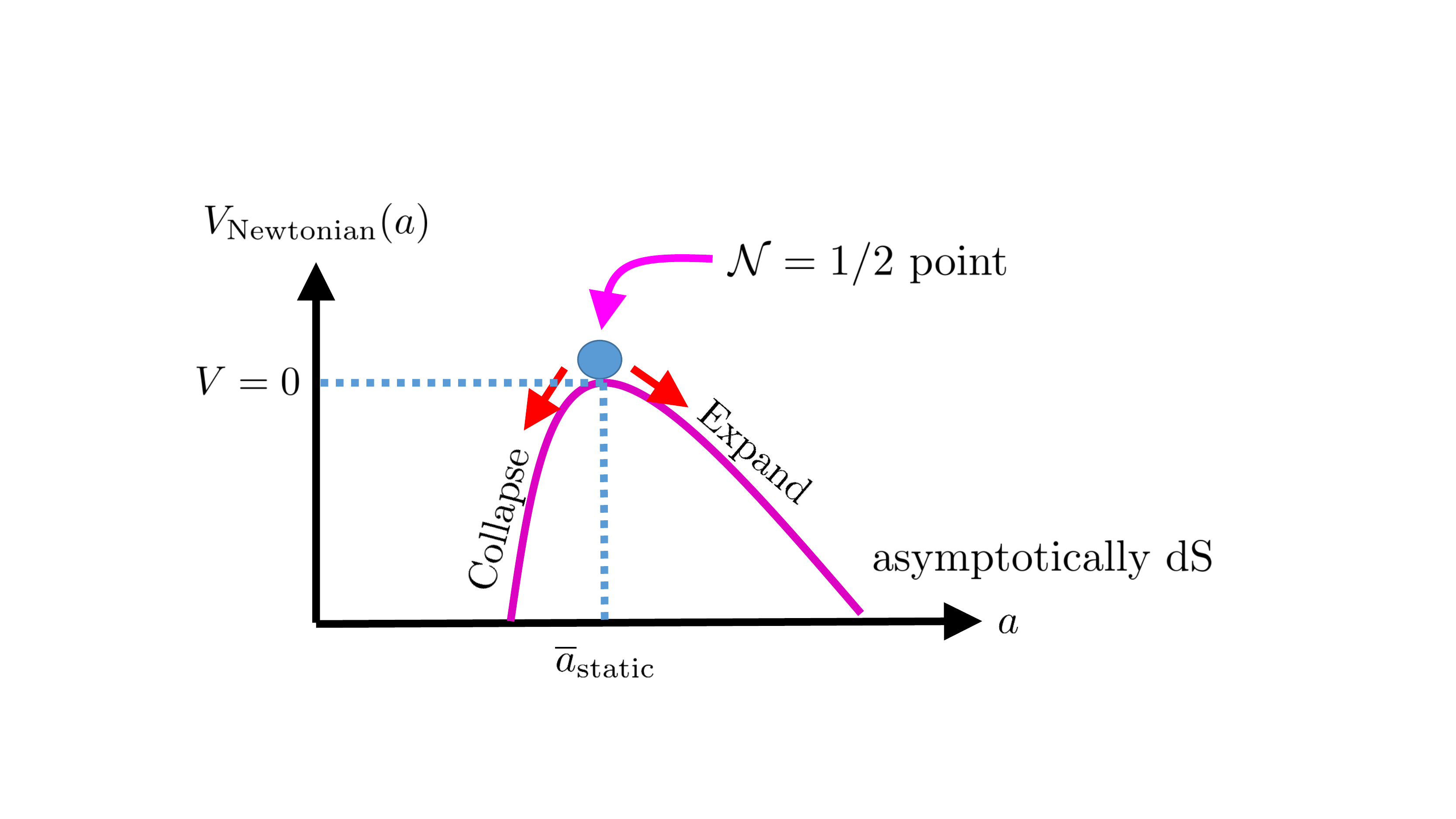}
\end{center}
\caption{
	Depiction of the instability in the static Universe solution obtained from a balancing between a stiff fluid, curvature and cosmological constant. Small perturbations lead to a subsequent rolling in the scale factor to either a collapsing or expanding phase.
We emphasize that no scalar potential is actually involved in this rolling motion.}
\label{fig:kingofthehill}
\end{figure}

Much as in the case of the Einstein static Universe, this model is unstable
against perturbations, and can either tip over into a collapsing or an
expanding phase (see figure \ref{fig:kingofthehill}).  The latter possibility
is a potentially promising avenue for building realistic models since it
asymptotically approaches de Sitter space at late times. Symmetry
considerations also suggest that the collapsing phase will eventually tunnel
out to the expanding branch \cite{Heckman:2018mxl}.

Of course, if this was all there was to the story, we could not use this as a
starting point for describing the observed Universe. For example, one must include an
inflationary epoch (or something else which generates the observed primordial density
perturbations), and then after inflation we expect to have an era of radiation
domination followed by matter domination, and finally cosmological
constant/dark energy domination.

One of the main results of this paper is that the model proposed in reference
\cite{Heckman:2018mxl} already contains the necessary ingredients for
realistic cosmology. This includes generating the correct scalar power
spectrum, and also includes a prediction that the equation of state for dark
energy is not an exact constant. These predictions
can be tested with current cosmological surveys such as Planck and
DES, and further constrained by future missions.

The scenario is as follows: although the unstable $\mathcal{N} = 1/2$
supersymmetric point contains no radiation (or matter), rolling from ``top of
the hill'' pushes the Universe away from a static vacuum, making tunneling events in
the flux integer $N$ possible. Each such tunneling event is accompanied by
some amount of radiation (more precisely, relativistic matter) which in turn
leads to a slow time dependence in the number of flux units $N(t)$. Because
$\Lambda(t)$ is a slowly varying function of time, we have an epoch of
inflation rather than pure de Sitter.  Even so, we stress that there is no
``compactification modulus'' playing the r\^{o}le of the inflaton, so in this
sense Swampland considerations have no direct bearing on our analysis.

\begin{figure}[t!]
\begin{center}
\includegraphics[trim={0cm 0cm 0cm 0cm},clip,scale=0.25]{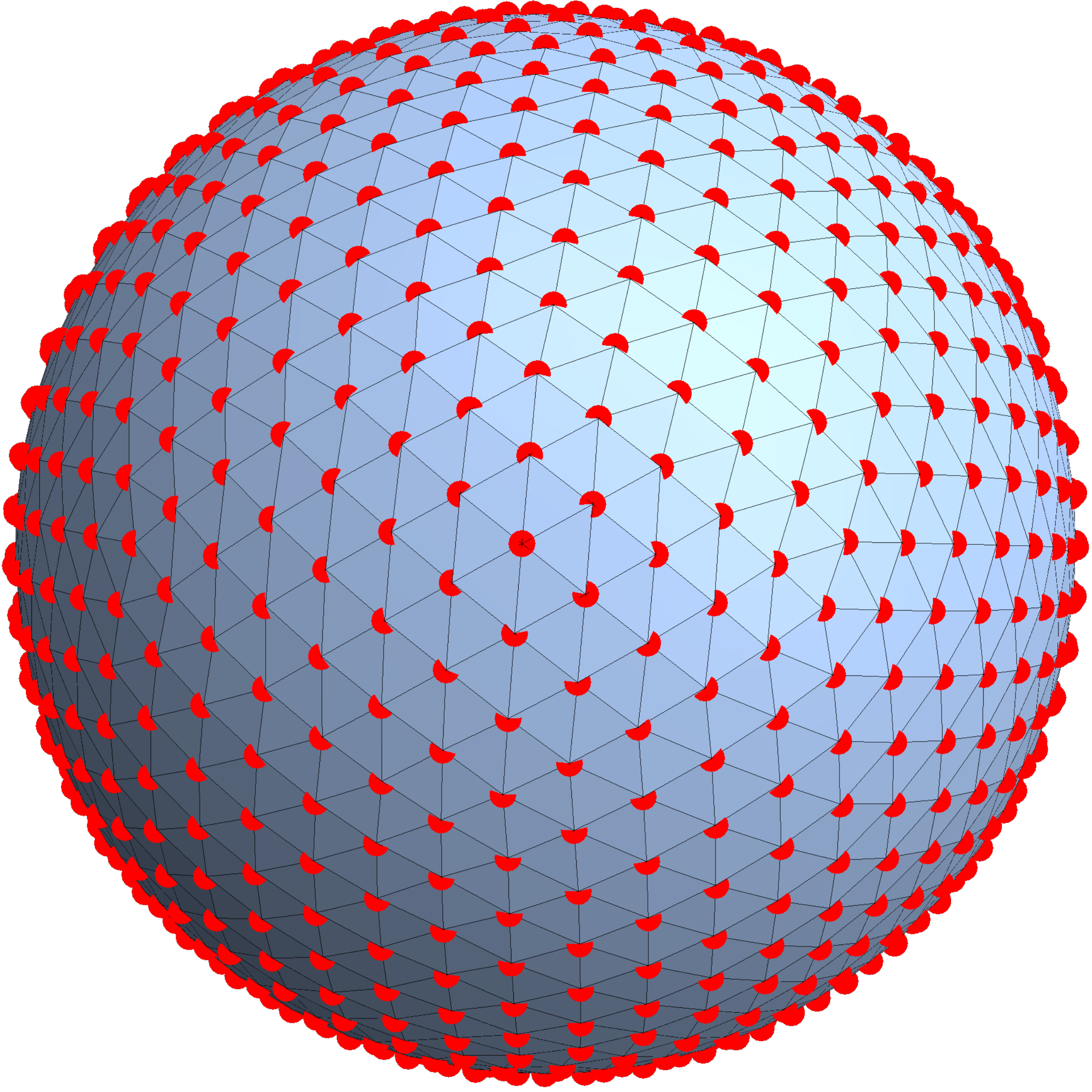}
\end{center}
\caption{
	Representation of a pixelated $S^3$. This leads to a cutoff on the angular momentum
of the physical system $l_{\text{max}} \sim N$ with $N$ the number of pixels.
In the stringy construction, each pixel originates from a five-brane
wrapped on a local five-cycle of the internal geometry.}
\label{fig:sphere}
\end{figure}

Indeed, the microscopic description is somewhat different from a 4D effective
field theory and fills in some conceptual points left open in reference
\cite{Heckman:2018mxl}.  The discretized parameter $N$ sets a fundamental
number of ``pixels'' for the actual spacetime. Each such pixel is associated with
a heavy particle which originates from a five-brane (that is, one time and
five spatial directions) which is wrapped over five internal directions of the
geometry. Locally, this five-brane can stay supported for a long time, but
eventually it can unwind in the internal geometry. This unwinding amounts to a
self-annihilation process which lowers the number of flux units, also emitting
radiation. We refer to this as the conversion of a pixel to a hole. We find that this is
entropically favored when the Universe is on the collapsing branch. Branes can
also wind up on metastable five-cycles. This is the conversion of a hole to a
pixel with final state radiation.  This process is entropically favored when
the Universe is on the expanding branch. See figure \ref{fig:sphere} for a depiction of a
pixelated $S^3$.

Now, although typically the rate of change in $N(t)$ is quite slow, every once
in a while there is a much larger jump in the flux parameter. Such rare events
generate a large amount of radiation and lead to a sudden jump to a radiation
dominated phase.  This is the analog of ``(pre-)reheating'' in standard
inflationary cosmology \cite{Kofman:1994rk}.  The pixel model of the Universe
also suggests that the transition to a Universe dominated by radiation sets
off a chain reaction in the breeding rate for pixels. So, even if the initial
jump in the number of flux quanta is relatively modest, it subsequently
triggers an exponentially large number of additional pixels to appear by the
end of the chain reaction,
\begin{equation}
N_{\text{reheat}} \sim N_{\text{ignite}} \times
e^{\Gamma_{\text{breed}}/T_{\ast}} \,,
\end{equation}
where $\Gamma_{\text{breed}} > 0$ is the pixel breeding rate. The
chain reaction terminates at a temperature $T_{\ast}$ at which the
system reaches thermal equilibrium between pixel/hole
conversion. After this point, the Universe expands enough
to dilute scattering between radiation and holes/pixels, and once this occurs
we move into the epoch of post-reheating and the start of Big Bang cosmology
with a slowly varying value for the flux $N(t)$. This in turn means that the
equation of state for dark energy is not exactly constant. The relative
strength of tensor to scalar perturbations depends on the time derivatives of
$N(t)$. The transition process then takes the form
\begin{equation}
N_{\text{inflation}}(t) \rightarrow N_{\text{ignite}} \rightarrow
N_{\text{reheat}} \rightarrow N_{\text{today}}(t) \,,
\end{equation}
with an exponential increase in the transition step $N_{\text{ignite}}
\rightarrow N_{\text{reheat}}$.

We also take some preliminary steps in determining the 1D matrix quantum
mechanics which governs these pixels. Our result is holographic in spirit and
leads to some intriguing features. For example, we find that there is an energetically preferred
fuzzy three-sphere geometry which retains all the isometries of the smooth
three-sphere. We interpret this as the backreaction of many such five-branes
as they dissolve into flux.

The pixel model also matches up well with the effective
thermodynamics experienced by observers moving in such a background. We find
that, much as in the case of M(atrix) theory black holes \cite{Banks:1996vh,
Banks:1997hz, Banks:1997tn}, the de Sitter temperature and entropy experienced
by an observer has an $N$-scaling
\begin{equation}
S_{\text{obs}} \sim N \quad \text{and} \quad T_{\text{obs}} \sim
\frac{1}{\sqrt{N}} \,.
\end{equation}
A detailed calculation of all the order one numerical pre-factors will certainly be
more challenging but this at least suggests a coherent starting point for
addressing such issues.

The rest of this paper is organized as follows. In section \ref{sec:COSMO} we
give a self-contained overview of the cosmology of the scenario, emphasizing
the observational consequences. The stringy underpinnings
are then developed in subsequent sections. In section
\ref{sec:INGREDIENTS} we summarize some of the recent progress made
in understanding F-theory on $\mathrm{Spin}(7)$ backgrounds. In section \ref{sec:GONEWITHTHEWIND}
we discuss how perturbations in the static solution lead to a winding/unwinding of five-branes
in the geometry and in section \ref{sec:JUMP} we analyze the energetics and entropics of the
flux changing as a function of time. Section \ref{sec:COLLAPSO}
presents a qualitative discussion of the collapsing branch of solutions, as well as
calculations suggesting that eventually the Universe will tunnel out to the expanding branch. Section
\ref{sec:REHEATING} discusses the out of equilibrium dynamics of pixels during
reheating and the subsequent increase in pixel number. In section \ref{sec:PIXEL}
we develop the quantum mechanical description of pixels and apply it to determine the scaling of the de Sitter entropy and temperature with pixel number $N$. We present our conclusions in section \ref{sec:CONC}. Some additional
technical material is deferred to the Appendices.

\section{Summary of the Cosmological Scenario} \label{sec:COSMO}

In this section we present an overview of the cosmological scenario.
Our starting point is a closed FRW Universe with metric
\begin{equation}
  \dd s_{\rm FRW}^2 = -\dd t^2 +{a^2(t)} \dd \Omega_{S^3}^2 \,.
\end{equation}
In particular, the string construction starts with a static Universe
with topology $\mathbb{R}_{\text{time}} \times S^3$, and with a balancing
between the curvature of the $S^3$, a stiff fluid with equation of state $P = \rho$,
and a cosmological constant term with equation of state $ P  = - \rho$. From the stringy construction, the zero point (vacuum) energies
cancel, so the value of the cosmological constant is protected against quantum fluctuations, and is therefore radiatively stable. For this reason, our model does not have a cosmological constant problem. The measured value is not fine-tuned and it does not make sense to ask why the observed value is one number rather than another, although, as we will see presently, a small number is natural from a UV-perspective.

Let us briefly expand on this point. The key observation in reference \cite{Heckman:2018mxl} is that in spacetimes of
signature $(2,2)$, supersymmetry permits us to organize supercharges into a real doublet (two real supercharges).
In signature $(3,1)$, supersymmetry instead organizes into complex doublets (Weyl fermions) so all finite energy excitations
are unprotected by supersymmetry. However, the ground state can be defined
by a process of analytic continuation from one signature to the other, much as how one defines
the Bunch--Davies vacuum for de Sitter space. Consequently, in the physical
spacetime of signature $(3,1)$, the ground state is protected by supersymmetry, but
all finite energy excitations experience broken supersymmetry. From a UV perspective, the main task
is to find string compactifications which would have preserved two supercharges in a 4D system
with signature $(2,2)$. This is where compactification of F-theory on $\mathrm{Spin}(7)$ backgrounds comes into the story.

The stringy model makes reference to a parameter $N$, the number of pixels tiling the spacetime, which in static solutions is fixed to be an integer. In terms of this
quantity, the energy densities of the stiff fluid and cosmological constant are
\begin{equation}
\rho_{\text{stiff}} = \frac{M_{\text{pl}}^2}{16 \pi^4}\frac{N^2 \ls^4 \kappa^3 }{a^6} \quad
\text{and}\quad  \rho_{\Lambda} = \frac{4 \pi^2 M_{\text{pl}}^2}{\ls^2 N} \,,
\end{equation}
where $\ls = 2 \pi \sqrt{\alpha'}$ is the string length scale, and $\kappa>0$ measures the curvature of the three-sphere.
This spacetime exhibits an instability, and so it inevitably either contracts or expands.

The stiff fluid dominates on the collapsing branch while the cosmological constant dominates on the expanding branch. In section
\ref{sec:COLLAPSO} we explain that even if the Universe ``gets unlucky'' and initially falls onto the collapsing branch, it can
tunnel out to the expanding branch. The reverse process is highly disfavored, so in what follows we just focus on the expanding branch.

Now, in the actual string
construction the value of $N$ should really be viewed as a time-dependent function. This is because
the ``pixels'' of the spacetime can either be created or destroyed and this process
is accompanied by some amount of radiation. For this reason, in what follows, we can
allow for the value of $N$ to have some time-dependence. Provided the rate of change in $N(t)$ is
small, the amount of radiation released in this transition will be sub-dominant compared to the cosmological constant. It can
happen, however, that a large jump in flux occurs. Such a rare event can allow radiation
to dominate over the other components. Our interpretation of this is that we have a period of slow-roll
inflation which comes to an end when such a rare decay process finally occurs.

With this as our starting point, the Friedmann equation is\footnote{Note that we have ignored the spatial curvature. This assumes we are starting after the Universe has inflated sufficiently that it is negligible. If this is not the case then one may expect signatures such as a correction to the normalization of the power spectrum \cite{Starobinsky:1996ek,Masso:2006gv}, and possibly a non-standard running of the spectral index \cite{Asgari:2015hna}.}
\begin{equation}\label{eq:friedearly}
  3H^2=\frac{8\pi^2}{\ls^2} \frac{1}{N(t)} \,,
\end{equation}
where $N(t)$ parameterizes our ignorance of the exact decay process. A crucial observation is that we have broken time-diffeomorphism invariance because $H$ is not constant and the spacetime is not exact de Sitter. This means we can write down an effective theory for perturbations about this spacetime by writing down all operators that are invariant under spatial-diffeomorphisms but not time-diffeomorphisms. Of course, this is just the effective field theory of inflation introduced in reference \cite{Cheung:2007st}:
\begin{equation}\label{eq:action1}
S=\int\dd^4x\sqrt{-g}\left[\frac{M_{\text{pl}}^2R}{2}+M_{\text{pl}}^2\dot{H}g^{00}-M_{\text{pl}}^2(3H^2+\dot{H})+\cdots\right]
\,.
\end{equation}
The form of the second two terms are fixed by requiring that the background evolution satisfies the Friedmann equations.
The dots involve contributions which are expected to be sub-leading to the main dynamics.\footnote{In the context of scalar field models, they correspond to operators that generate appreciable non-Gaussianities such as those that arise in theories with non-canonical kinetic terms, as in DBI inflation \cite{Silverstein:2003hf} for example.}
The relevant terms correspond precisely to single-field slow-roll inflation. To see this, consider the action
\begin{equation}
S=\int\dd^4x\sqrt{-g}\left[-\frac{1}{2}(\partial\phi)^2-V(\phi)\right]\rightarrow\int\dd^4x\sqrt{-g}\left[-\frac{\dot{\phi}(t)^2}{2}g^{00}-V(\phi(t))\right]
\,.
\end{equation}
In Appendix \ref{sec:inflation} we review how one can construct the primordial power spectrum and CMB observables from this formalism. The result is $n_s-1=-2\ve-\eta$ where the expressions for $\ve$ and $\eta$ are \cite{Leach:2002ar,Akrami:2018odb}
\begin{equation}\label{eq:epseta}
\varepsilon \equiv-\frac{\dot{H}}{H^2} \quad \text{and} \quad \eta \equiv
\frac{\dot{\ve}}{H\ve}=\frac{\ddot{H}}{H\dot{H}}-2\frac{\dot{H}}{H^2} \,.
\end{equation}

To summarize, the fact that we have broken time-diffeomorphism invariance by
having a changing cosmological constant implies that the universe is quasi-de
Sitter, and we can identify the Goldstone boson of this broken symmetry with the perturbations of an inflaton. It is interesting to note that one could reconstruct an effective inflaton potential (or at least, its first and second derivatives) from knowledge of the observed power spectrum that would give identical predictions to our pixel model. Such a formalism would require two initial conditions, $\phi_i$ and $\dot{\phi}_i$. Some of the criticism of inflation is that the initial conditions are fine-tuned. Whilst an observer reconstructing the potential sees the initial conditions as fine-tuned, such a tuning is natural from the point of view of the pixel construction since the entire equivalent history of $\phi(t)$ is fixed by $N(t)$, which is in turn set by the rate of winding up of 5-branes. Furthermore, the fact that we do not have a fundamental scalar in our description avoids other criticisms connected with the issue of eternal inflation and the homogeneity of the initial conditions on scales larger than the initial horizon.

\subsection{Observables}

In this subsection we use the formalism developed above to extract the main observational signatures of this class of models.
This includes the spectral index for scalar perturbations as well as a parametrization of the equation of state for
dark energy in the present framework.

\subsubsection{Inflation and the Calculation of the Spectral Index}

To find the spectral index we simply need to calculate the slow roll parameters $\ve$ and $\eta$. This can be done as follows.
Using equation \eqref{eq:friedearly} one has
\begin{equation}\label{eq:Hs}
H=\sqrt{\frac{8\pi^2}{3\ls^2}}\frac{1}{\sqrt{N_{\text{inf}}}} \, , \quad
\dot{H}=-\sqrt{\frac{8\pi^2}{3\ls^2}}\frac{{\dNi}}{2 N_{\text{inf}}^{3/2}} \,
, \quad\ddot{H}=\sqrt{\frac{8\pi^2}{3\ls^2}}
\left(\frac{3}{4}\frac{{\dNi}^2}{N_{\text{inf}}^{5/2}} - \frac{1}{2}
\frac{\ddNi}{N_{\text{inf}}^{3/2}}\right) \,,
\end{equation}
where $N_{\text{inf}}$ is the value of $N$ during inflation i.e. before the rare jump that reheats the Universe. These can be used in the definitions \eqref{eq:epseta} to find
\begin{equation}\label{eq:SRparams}
\ve=\sqrt{\frac{3 l_{s}^2}{32 \pi^2}}\frac{\dNi}{\sqrt{N_{\text{inf}}}} \,
,\quad\eta = -\left(1-2\frac{N_{\text{inf}}{\ddNi}}{{\dNi}^2}\right)\ve \,.
\end{equation}
The spectral index is then
\begin{equation}\label{eq:specind}
  n_s-1=-2\ve-\eta=-\left(1+2\frac{N_{\text{inf}}\ddNi}{{\dNi}^2}\right)\ve \,.
\end{equation}
Since this is single field slow roll, the tensor to scalar ratio is
\begin{equation}
  r=16\ve= \sqrt{\frac{8 \, \ls^2}{\pi^2}}\frac{\dNi}{\sqrt{N_{\text{inf}}}}
  \,.
\end{equation}
Finally,  we see that the amplitude of the power spectrum (defined in Appendix \ref{sec:inflation}) is given by
\begin{equation}\label{eq:amp}
A_s=\frac{\pi H_\ast^2}{\ve_\ast M_{\text{pl}}^2}=2\pi\left(\frac{8\pi^2}{3}\right)^{3/2} \frac{M_{\text{pl}}}{\sqrt{N_{\text{inf}}}\dNi} \, ,
\end{equation}
where subscript asterisks denote quantities evaluated at horizon crossing ($k=aH$). In the above we have also
taken the rough approximation $\ls \sim M_{\text{pl}}^{-1}$.

We can use the Planck constraints on slow-roll inflation \cite{Akrami:2018odb} to place bounds on the unknown time-dependence of $\Ni(t)$ and its derivatives\footnote{Here we have used the constraints $\ve<0.0063$ (95\% C.L.) and $\eta=0.030$ (68\% C.L.) with $n_s-1=-0.0351$, $r<0.069$, and $A_s=2\times10^{-9}$ \cite{Akrami:2018odb}.} using the formulae in equation \eqref{eq:SRparams} and \eqref{eq:specind}. From the Planck bound on $\ve$ and $n_s-1$ we find
\begin{equation}\label{eq:infboubds1}
  \frac{\ls\dNi}{\sqrt{\Ni}}<0.07 \, , \quad \frac{\ddNi\Ni}{\dNi^2}>3 \,.
\end{equation}

In single field slow-roll models, one needs a measurement of $\ve$ (e.g. by measuring the scalar-to-tensor ratio) to pin down the scale of inflation, and the same is true for our construction, however we can learn about some fundamental quantities during inflation from the Planck measurement of $A_s$ alone. Using the relations \eqref{eq:Hs} one has
\begin{equation}
  \frac{\dNi\sqrt{\Ni}}{M_{\text{pl}}}=4\times10^{11} \,.
\end{equation}
Using equation \eqref{eq:infboubds1}, one can eliminate $\dot{N}$ to find
\begin{equation}\label{eqn:Ninf}
  \Ni>6\times10^{12} \,.
\end{equation}
The rate of flux wind-up during inflation (or pixel creation from the 4D perspective) is then
\begin{equation}
  \Gamma_X \big\vert_{\text{inf}}=\frac{\dNi}{\Ni}<3\times10^{-8}M_{\text{pl}} \,,
\end{equation}
i.e. it is sub-Planckian. Since the decay rate implicitly depends on details of the geometry (which are themselves $N$-dependent) this
will generically not be the same as the present day value of $\Gamma_X$.

We can also use this bound on $N_{\text{inf}}$ to determine an upper
bound on the Hubble parameter during inflation. Returning to
our expression for the Hubble parameter, we have
\begin{equation}
H_{\text{inf}} = \sqrt{\frac{8 \pi^2}{3 \ell_s^2}}
\frac{1}{\sqrt{N_{\text{inf}}}} \lesssim  2 \times 10^{12} \, \text{GeV} \,.
\end{equation}
This also sets a crude limit for the temperature at the start of reheating, which we call the ``ignition temperature:''
\begin{equation}\label{Tignite}
T_{\text{ignite}} \sim \sqrt{ H_{\text{inf}} M_{\text{pl}}} \lesssim 10^{15} \,
\text{GeV} \,,
\end{equation}
which is no different than standard expectations from single field slow roll inflation.

\subsubsection{Absence of Isocurvature Perturbations}

The three-form flux in our model is magnetically dual to the gradient of an
axion ($H_3 = \ast F_1$). The fact that its energy redshifts like that of a
stiff fluid implies that it is massless and so one may be concerned that it
acquires large isocurvature perturbations during the inflationary phase that
are incompatible with CMB data. As discussed in section
\ref{sec:GONEWITHTHEWIND}, the axion couples to gauge fields\footnote{The
axion in our construction arises from a closed string mode and therefore is
expected to couple to every sector.}, which in turn give rise to a potential
via non-perturbative (instanton) effects in an analogous manner to the
mechanism by which to the QCD axion gains a mass. Since many sectors are
expected to be strongly coupled, the confinement scale, and therefore the
axion mass can in principle be quite large. Isocurvature perturbations
will be highly suppressed when
\begin{equation}
  m_{\text{axion}} \gtrsim H_{\text{inf}} \,.
\end{equation}
Since we have already obtained a rough bound on $H_{\text{inf}} \lesssim
10^{12}$ GeV, we conclude that this can indeed be satisfied in a number of
stringy scenarios. Let us also note that there is an extensive model building
literature centered on ways to evade isocurvature bounds for low mass axions.
See e.g. \cite{Graham:2018jyp} and references therein for recent
discussion on this point.

\subsubsection{Cutoff on the Angular Power Spectrum}

A novel and interesting prediction of our construction is that the Universe is
a fuzzy three-sphere, by which we mean that the geometry is actually composed
of $N$ pixels so that on large scales the spacetime appears as a smooth
three-sphere. The cutoff goes like
\begin{equation}
  l_{\text{max}} \sim N \,,
\end{equation}
and, as we have already discussed in equation (\ref{eqn:Ninf}), this is a rather high number, at
least on the order of $10^{12}$. Current limits from CMB experiments do not
impose significant constraints. For example, Planck probes the CMB on scales
$l_{\rm res}=2500$ in the TT-spectrum \cite{Aghanim:2018eyx} (less in the EE-
and cross spectra), which correspond to $0.05\degree<\theta<0.1\degree$.
Ground based experiments that cover a small patch of the sky can reach even
smaller scales (larger values of $l$). In particular, the South Pole Telescope
(SPT) has probed $l_{\rm res}=8000$ in the EE-power spectrum
\cite{Henning:2017nuy}. The Atacama Cosmology Telescope (ACT) reaches a
similar resolution \cite{Louis:2016ahn}. Future CMB-S4 experiments may improve
this bound by an order of magnitude. It is also possible that optical lensing
surveys of galaxies could produce a more significant constraint, though again,
it is unclear whether the extreme resolution of $l_{\text{max}} \sim 10^{12}$  would be
reached.

\subsubsection{Thermal History}

Inflation ends after a large jump where $N_{\text{inf}}\rightarrow
N_{\text{inf}}+k=N_{\rm ignite}$ with $k>1$. This decay is suppressed relative
to jumps with $k=1$ (see section \ref{sec:GONEWITHTHEWIND}) and so it is
expected that inflation lasts a long time. For this reason, $k$ is expected to
be the smallest number such that reheating occurs. In section \ref{sec:JUMP},
we show that after a rare decay the Universe is radiation-dominated with a
large amount of radiation given by
\begin{equation}
  \frac{\rho_\text{rad}}{\rho_\Lambda}=\frac{k}{N_{\text{inf}}} \,.
\end{equation}
As we discuss in section \ref{sec:REHEATING} there is then a further
exponential increase from $N_{\text{ignite}}$, the value at the start of
reheating, to the final value $N_{\text{reheat}}$.  We present an idea along
these lines in section \ref{sec:REHEATING} where we argue that in the process
of exiting inflation, the emission of radiation by the birth of pixels in our
spacetime triggers a chain reaction,
\begin{equation}
N_{\text{reheat}} \sim N_{\text{ignite}} \, e^{\Gamma_{\text{breed}}/T_{\ast}}
\,,
\end{equation}
where $T_{\ast}$ is the temperature at which the chain reaction terminates and
$\Gamma_{\text{breed}}$ is the pixel breeding rate. So, even if
$N_{\text{ignite}}$ is a modestly small number, it can rapidly generate a
final large value for $N_{\rm reheat}$. After this chain reaction comes to an
end, we get a zero-temperature cosmological constant,
\begin{equation}
  \Lambda=\frac{8\pi^2}{\ls^2} \frac{1}{N_{\text{reheat}}}\sim 3 H_0^2 \,,
\end{equation}
where the order of magnitude is fixed by the measurement of the cosmological
constant today. This is justified since the cosmological constant is both
UV-insensitive and radiatively stable.

An important comment is that in the limit of zero temperature the
system recovers ``$\mathcal{N} = 1/2$'' supersymmetry, i.e. we expect the
ground state to be protected against zero point energy contributions. This in
turn sets the relative energy scales for all phase transitions at higher
temperatures, since we require the $T = 0$ limit to preserve this property.

Now the magnitude of the cosmological constant driving the expansion of the
Universe after reheating is larger than the zero temperature value because of
contributions from phase transitions. This is because the Universe has jumped
into a high-temperature phase where the temperature may be above either the
QCD or electroweak phase transition (EWPT). One then has
\begin{equation}
\frac{\rho_\text{rad}}{\rho_\Lambda}=\frac{\pi^2}{30}g_\ast\left(\frac{T_{\rm
reheat}}{E_{\rm PT}}\right)^4 \,,
\end{equation}
where $E_{\rm PT}$ is the energy scale of the phase transition.  If the
reheating temperature is above the EWPT we have $E_{\rm PT}=200$ GeV. If
instead the reheating temperature is lower than this but above that of the QCD
phase transition one has $E_{\rm PT}=100$ MeV. Below $100$ MeV, $E_{\rm PT}$
is simply the zero-temperature vacuum energy (meV), although one cannot reheat
below $4$ MeV and still have big bang nucleosynthesis \cite{Hannestad:2004px}.
No matter the reheating temperature, our model does not suffer from any tuning
because the phase transition occurs when $T=E_{\rm PT}$, at which point the
ratio of the radiation and cosmological constant densities is
$\frac{\pi^2}{30}g_\ast>1$ so the universe is still radiation dominated and
the large CC vanishes before it can dominate.

We can now fix the parameter $N_{\rm reheat}$ to its observed value today, in
terms of the string scale.  We have\footnote{Note that after the chain
reaction ends the value of $N$ winds up by one pixel at a time so that $N_{\rm
reheat}\sim N_{\rm today}$.}
\begin{equation}
\Lambda = 4.33 \times 10^{-84} \, \text{GeV}^2 = \frac{8 \pi^2 M_{\text{pl}}^2}{N_{\rm reheat}} \, ,
\end{equation}
where again we have set $\ls \sim M_{\text{pl}}^{-1}$. We then have
\begin{equation}
  N_{\rm reheat} \sim 10^{120} \,.
\end{equation}
A priori, there is no bound on the value of $N_{\rm reheat}$ and in
fact we need $N_{\rm reheat}$ large to justify some of our approximations.

Since we have fixed the overall zero temperature constants appearing in the
free energy of all phase transitions, we also expect that in regions of the
Universe which are still at high temperature, the local vacuum energy density
and thus the local value of the cosmological constant to be higher. This will
occur, for example, in the interior of neutron stars, and it was proposed in
references \cite{Bellazzini:2015wva,Csaki:2018fls, Han:2018mtj} that the
effects of this could be detected via gravitational waves generated by a binary
merger event.

\subsubsection{Dark Energy}

The fact that we have a non-zero decay rate during the early universe suggests
that the same is true today. Including matter, the Friedmann equation is
\begin{equation}
  \label{eq:Friedm1}3H^2 M_{\text{pl}}^{2}=\rho_m+\rho_\Lambda \,,
\end{equation}
where
\begin{equation}
\rho_\Lambda(t)= \frac{4\pi^2 M_{\text{pl}}^{2}}{\ls^2N(t)} \,.
\end{equation}
In order for the Friedmann equation to be consistent with the acceleration equation,
the cosmological constant must satisfy the continuity equation
\begin{equation}
  \dot{\rho}_\Lambda+3H(\rho_\Lambda+P_\Lambda)=0 \,,
\end{equation}
which implies that there is a pressure
\begin{equation}
  P_\Lambda=-\rho_\Lambda\left(1-\frac{\dot{N}}{3HN}\right) \,,
\end{equation}
so that the equation of state deviates from a cosmological constant. In fact, one has
\begin{equation}\label{eq:w1}
  w=-1+\frac{\dot{N}}{3HN} \,.
\end{equation}
We can estimate this as follows. Writing the Friedmann equation as
\begin{equation}
  H^2=\frac{\Omega_{m,0}H_0^2}{a^3}+\frac{\Lambda}{3} \,,
\end{equation}
and Taylor expanding $a$ to first order we find
\begin{equation}
  H(a)=H_0\left[1+\frac{3}{2}\Omega_{m,0}(1-a)+ O((1-a)^2)\right] \,,
\end{equation}
which can be substituted into \eqref{eq:w1} to yield
\begin{equation}\label{eq:W0Wa}
  w(a)=-1+\frac{\dot{N}}{3H_0N}-\frac12\Omega_{m,0}\frac{\dot{N}}{H_0N}(1-a) \,.
\end{equation}
Our model clearly fits into the $w_0$--$w_a$ parametrization, $w=w_0+w_a(1-a)$,
which is a common way to characterize the dark energy equation of state \cite{Linder:2005ne}.
The construction predicts the relation
\begin{equation}\label{eq:w0wa}
  w_a=-\frac{3}{2}\Omega_{m,0}(1+w_0) \,.
\end{equation}

We can place constraints on the value of $N$ today using observations of dark
energy and our predictions for $w_0$ and $w_a$ given in equation
\eqref{eq:W0Wa}. Using the DES Y1 results (which are consistent with Planck)
\cite{Abbott:2018xao} we have $w_0=-0.95^{+0.09}_{-0.08}$ and
$w_a=-0.28^{+0.37}_{-0.48}$ at the 68\% confidence level. Denoting the number
of pixels today (which should be similar to the number at reheating) by $\Nd$($=N_{\rm today}$),
this implies that $-0.09<\dNd/(H_0\Nd)<0.42$, which implies $-0.02<w_a<0.07$
at the 1$\sigma$ level, so our predictions are currently consistent with
cosmological data, although a cosmological fitting using equation
\eqref{eq:w0wa} would probably yield more robust information. Future data
releases would constrain $\dNd/(H_0\Nd)$ further, and upcoming surveys such as
Euclid and LSST would provide even tighter bounds.

Note that we have not assumed any correlation between the winding-up (pixelation) rate today and the rate during inflation. Indeed, the current bounds imply that the rate today is
\begin{equation}
  \frac{\Gamma_X}{M_{\text{pl}}} \bigg\vert _{\text{today}} <4\times10^{-61} \,,
\end{equation}
far slower than during inflation.

\subsection{The Rest of the Universe}

There is, of course, an extensive list of additional requirements necessary to
build a fully realistic model.  Here we briefly discuss some of these issues.

\subsubsection{Dark Matter}

It is also natural to discuss the other dark component of our Universe: dark
matter. Since our model contains an axion it is of course tempting to ask
whether this can play the r\^{o}le of dark matter. As alluded to above, and
discussed in detail in section \ref{sec:GONEWITHTHEWIND}, we expect the mass
of the axion to be large due to its coupling to strongly coupled gauge sectors
with high confinement scales. For this reason, our axion is not a good dark
matter candidate since it will rapidly decay to the gauge fields leaving a
negligible relic density. Of course, string theory contains a profusion of
hidden-sector fields, any one of which could contain a stable dark matter
candidate. For some specific examples in the context of F-theory based
constructions, see e.g.~references \cite{Heckman:2008jy, Heckman:2009mn,
Heckman:2011sw, DelZotto:2016fju}.

\subsubsection{Superpartners}

Another intriguing feature of the model presented in reference
\cite{Heckman:2018mxl} is that one can also roughly estimate the mass scale
for superpartners. As explained there, this is specified by the geometric mean
set by the UV cutoff (the 4D Planck scale) and the IR cutoff, as set by
$\rho_{\text{vac}} \sim M_{\rm IR}^4$:
\begin{equation}\label{splitto}
  \Delta m_{\rm SUSY} \sim \sqrt{M_{\rm IR} M_{\rm UV}} \,,
\end{equation}
or in terms of the parameter $N_{\rm reheat}$ and the 4D Planck scale, we expect
\begin{equation}
  \Delta m_{\rm 4D} \sim \frac{1}{N_{\rm reheat}^{1/8}} M_{\text{pl}} \,.
\end{equation}
Strictly speaking, this formula only applies when the $S^3$ of the geometry is
static, i.e. for vanishing Hubble parameter. However, the expansion rate
generates a subleading contribution to the mass splitting of order $\sqrt{H_0
M_{\text{UV}}} \ll \sqrt{M_{\rm IR} M_{\rm UV}}$.  Since we have also fixed
the present day value of $N_{\rm reheat} \sim 10^{120}$, we are led to the
same conclusion as in reference \cite{Heckman:2018mxl} that the superpartner
masses are on the order of $O(10 - 100)$ TeV, and the same caveats apply,
namely this is to be treated as a crude order of magnitude estimate at best.

We remark that equation (\ref{splitto}) is really a byproduct of
self-consistency conditions, but we do not at present have a first principles
derivation of this $N$ scaling. We expect that the pixel description we
develop in later sections will provide a more systematic derivation of these
scaling relations.

Supersymmetry breaking also affects various details of moduli. Simply because
the superpartner mass splittings can easily be on the order of $100$ TeV, we expect that
some of the typical issues which arise in string constructions (such as late
decaying scalars spoiling BBN) are automatically dealt with. See e.g.~reference
\cite{Kane:2015jia} for further discussion on this point.  A more delicate
issue concerns the potential appearance of moduli with a runaway direction in
the potential energy density.  Provided these moduli decay sufficiently early,
it will likely simply lead to a non-thermal history for part of the early
Universe. An explicit scenario along these lines involving the saxion was
presented in the early F-theory GUT literature, e.g.~reference
\cite{Heckman:2008jy}.

\subsubsection{Future Rare Events}

As the Universe continues to expand, the contribution from matter is expected
to again dilute away, leaving an era of dark energy with a slow time variation
in the equation of state. On sufficiently long time scales, we can expect a
future rare event to occur in which radiation again comes to dominate.
Observe that each such jump leads to a corresponding increase in the number
of pixels. As the number of pixels increases, there is an increase in time
between each such rare event. A priori, however, there seems to be no issue
with the system continuing in this way indefinitely.

\section{The Spin(7) Universe} \label{sec:INGREDIENTS}

Having sketched the main qualitative features of our scenario, we now turn to
a more detailed discussion of how these ingredients actually arise. In this
section we introduce the minimal features required to understand the
cosmological elements of F-theory on $\mathrm{Spin}(7)$ backgrounds. We refer
the interested reader to reference \cite{Heckman:2018mxl} for additional
discussion on the details of this proposal. From a 4D perspective, the main
idea is to consider spacetimes of the form $\mathbb{R}_{\text{time}} \times
S^3$. A note on conventions: starting from this section we will employ
units where the curvature of the $S^3$ is set to 1, that is $\kappa=1$.
While this may not be the standard choice in the cosmology literature it greatly
simplifies many equations and it is compatible with the conventions taken in
\cite{Heckman:2018mxl}. To reinstate curvature in the following it is
sufficient to replace everywhere the scale factor $a(t)$ with
$a(t)/\sqrt{\kappa}$.

An important phenomenological feature of such $\mathrm{Spin}(7)$ backgrounds
(as motivated by the earlier work \cite{Witten:1994cga, Witten:1995rz,
Vafa:1996xn}) is that the ground state of the system is annihilated by two
real supercharges but all finite energy excitations do not enjoy a
supersymmetric mass degeneracy.  A crude estimate for the mass of the
superpartners was obtained in reference \cite{Heckman:2018mxl}:
\begin{equation}
  \Delta M_{\text{SUSY}} \sim \sqrt{M_{\rm IR} M_{\rm UV}} \,,
\end{equation}
where $M_{\rm UV}$ is the high energy cutoff (the Planck scale) and $M_{\rm
IR}$ is the IR cutoff as specified by the vacuum energy density. In terms of
the radius of the $S^3$ and the UV cutoff, we have
\begin{equation}
  M_{\rm IR}^{4} \sim \frac{M_{\rm UV}^2}{a_{\text{today}}^2} \,.
\end{equation}
Strictly speaking this formula assumes the $S^3$ is expanding slowly, i.e. it is
``close enough'' to a static solution. There will be corrections to
this mass splitting, as well as to the vacuum energy density, as set by the
scale $\sqrt{H_0 M_{\rm UV}} \sim M_{\rm IR}$, with $H_0$ the Hubble
parameter. Observe, however, that this is an extremely small correction to the
superpartner masses, and is of the same order of magnitude as the vacuum
energy density. In this sense, the small value of the cosmological constant is
technically natural.

In F-theory \cite{Vafa:1996xn,Morrison:1996na,Morrison:1996pp}, the 10D spacetime of string theory is supplemented by two
additional directions given by a two torus, which we write as $T^2$. The shape
of this $T^2$ dictates the strength of the string coupling, and can in
principle be position dependent. In F-theory, physical degrees of freedom
still only propagate in 10D.

Supersymmetric vacua in F-theory are obtained by working with internal
geometries which preserve a covariantly constant spinor, namely special
holonomy manifolds.  The case of an eight-manifold with $\mathrm{Spin}(7)$
holonomy rather than the generic $\mathrm{Spin}(8)$ holonomy yields a solution
which in 4D spacetime signature $(2,2)$ would preserve two real supercharges. In the physical
$(3,1)$ signature, supersymmetry is absent for all finite energy excitations, but the ground state (being defined by analytic continuation
from one signature to the next) is still protected.

Additionally, to consistently solve the Einstein field equations we need to consider 4D
spacetimes with topology $\mathbb{R}_{\text{time}} \times S^3$.  The $S^3$
does not collapse because it is supported by a Neveu--Schwarz three-form flux
$H_3$ with $N$ flux units,
\begin{equation}
  \frac{1}{ \ls^2} \int_{S^3} H_3 =  N \,,
\end{equation}
with $N$ an integer. This three-form flux is the magnetic dual to a one-form
field strength
\begin{equation}\label{axionflux}
  F_{1} = \ast_{\rm 4D} H_{3} \,.
\end{equation}
Expanding around a background field configuration, we can introduce a small
fluctuation which we respectively identify with an axion $\varphi$ and a
two-form potential $B_2$ such that $d \varphi \sim F_1$ and $d B_2 \sim H_3$.

The presence of this three-form flux is necessary because on a
$\mathrm{Spin}(7)$ background we expect a non-holomorphic profile for the IIB
axio-dilaton. This leads to a stronger gradient profile compared with the case
of elliptically fibered Calabi--Yau manifolds, which in turn requires
additional supergravity fluxes be present to satisfy the Einstein field
equations.  The setup bears an interesting resemblance to recent
``clockwork/linear dilaton'' constructions in the phenomenology literature
(see e.g. \cite{Giudice:2016yja, Giudice:2017fmj}) which leads to a dilute 4D
Newton's constant, even though the internal directions could be quite large.
In this sense, the $\mathrm{Spin}(7)$ construction is a stringy implementation
of these ideas, much as reference \cite{Verlinde:1999fy} was a stringy
proposal for implementing warped extra dimensions \cite{Randall:1999ee,
Randall:1999vf}. Let us note, however, that we still expect the Kaluza--Klein
length scale to be quite small relative to the size of our spacetime $S^3$.
The reason is that because the $S^3$ with $H$-flux is effectively flat (with
respect to the twisted spin connection), there appears to be no correlation
between the volume of the extra dimensions and the size of the $S^3$. This is
rather different from standard Freund--Rubin compactification where the $AdS$
radius and extra dimension radius are often correlated. The sign and size of the cosmological
constant are controlled by the gradient of the dilaton in the internal directions:
\begin{equation}\label{naboo}
\Lambda \sim (\overrightarrow{\nabla} \phi_{\text{dil}} )^2 \sim \frac{1}{N}.
\end{equation}
The sign is fixed because this term descends from the kinetic term of a 10D field. The magnitude is fixed
by the standard profile of a fluxed $S^3$. We note that in addition to this we
expect to be able to tune independently the internal volume before the
introduction of non-holomorphic piece of the axio-dilaton  (specifically the
length of the throat region of the linear dilaton background) thus effectively
decoupling the size of the internal space from the volume of the $S^3$.

Another perspective which helps to establish this claim and
will prove especially helpful in the present work is
to view the $\mathrm{Spin}(7)$ background with ``$\mathcal{N} = 1/2$
supersymmetry'' as the gravitationally backreacted spacetime obtained from
five-branes wrapped on five-cycles of an $\mathcal{N} = 1$ F-theory
background.  This is obtained from a Calabi--Yau fourfold with holonomy
$\mathrm{SU}(4) \subset \mathrm{Spin}(8)$.  Wrapping five-branes on local
five-cycles in this geometry yields a collection of point particles which
breaks half the supersymmetry, and also breaks Lorentz symmetry. The latter
issue can be addressed by suitably distributing these five-branes over the 4D
spacetime, and the backreacted limit produces an $S^3$ tiled (in the large $N$
limit this is known as ``smearing'') by these point particles. In section
\ref{sec:PIXEL} we discuss in more detail the quantum mechanical matrix model
which governs these pixels. This construction also makes it clear that most model building considerations
used to generate MSSM-like vacua in F-theory (see e.g. \cite{Beasley:2008dc,
Donagi:2008ca} and references \cite{Heckman:2010bq,Weigand:2010wm} for early
reviews) can also be carried over without much change.

From the perspective of the 4D spacetime, we are dealing with a particular
static FRW cosmology with topology $\mathbb{R}_{\text{time}} \times S^3$ in
which we have different contributions to the energy density, as dictated by
the equation of state $P = w \rho$. We have a positive cosmological constant
(with $w = -1$), a stiff fluid (with $w = +1$) and curvature all balancing
against one another. Writing the metric as
\begin{equation}
  \dd s^2_{\rm FRW} = -\dd t^2 +  { a(t)^2}\dd \Omega^2_{S^3} \,,
\end{equation}
The Hubble equation reads
\begin{equation}\label{Hubba}
H^2 = -\frac{1}{a^2} + \frac{1}{3}\frac{N^2 \ls^4  }{(2 \pi)^4 a^6} +
\frac{\Lambda}{3} \,,
\end{equation}
with Hubble parameter given by $H = \dot{a}/a$ as usual. The supersymmetric point
corresponds to the special case where $H = 0$, and the cosmological constant
scales as $1/N$:
\begin{equation}
  \Lambda = \frac{8 \pi^2}{\ls^2 N} \,,
\end{equation}
and the size of the $S^3$ at the ``top of the hill'' is controlled by
\begin{equation}
\overline{a}^2 = \frac{N \ls^2 }{(2 \pi)^2}.
\end{equation}
The various energy densities scale with $N$ as
\begin{equation}
  \rho_{\text{stiff}}(N) = \rho_{\text{stiff}}^0(1) \frac{N^2 }{a^6} \,, \quad
  \rho_{\Lambda}(N) = \rho_{\Lambda}^0(1)
  \frac{1}{N} \,, \quad \rho_{\text{curvature}}(N) =
  \rho_{\text{curvature}}^0(1) \frac{1}{a^2} \,,
\end{equation}
where the numerical coefficients are
\begin{equation}
  \rho_{\text{stiff}}^0(1) = \frac{M_{\text{pl}^2}\ls^4}{16 \pi^4} \,,\quad \rho_{\Lambda}^0(1) =
  \frac{8 \pi^2 M_{\text{pl}}^2}{\ls^2} \,,\quad \rho_{\text{curvature}}^0(1) =
  -3 M_{\text{pl}}^2 \,.
\end{equation}
For a depiction of this instability, and the subsequent rolling down a hill in
terms of an effective Newtonian potential, see figure \ref{fig:kingofthehill}.

Much as in the case of the Einstein static Universe (where we have ordinary
matter instead of a stiff fluid) this system is unstable against perturbations
and will either collapse or expand. To see this, consider equation
(\ref{Hubba}). This specifies a first integral for the classical dynamics of
the scale factor.  As a sum of kinetic and potential energy, we see that the
kinetic energy is formally negative, and so as a classical mechanics problem
the scale factor rolls off the top of the hill. Part of our aim will be to
study what happens to the resulting cosmology as we move away from this
special supersymmetric configuration.

\section{Winding and Unwinding Pixels} \label{sec:GONEWITHTHEWIND}

So far, we have treated the flux in our system as a fixed quantity. As we now
explain, once the Universe moves off the static solution, we should expect
there to be jumps in the number of flux units.

\begin{figure}[t!]
\begin{center}
\includegraphics[width = .99\hsize]{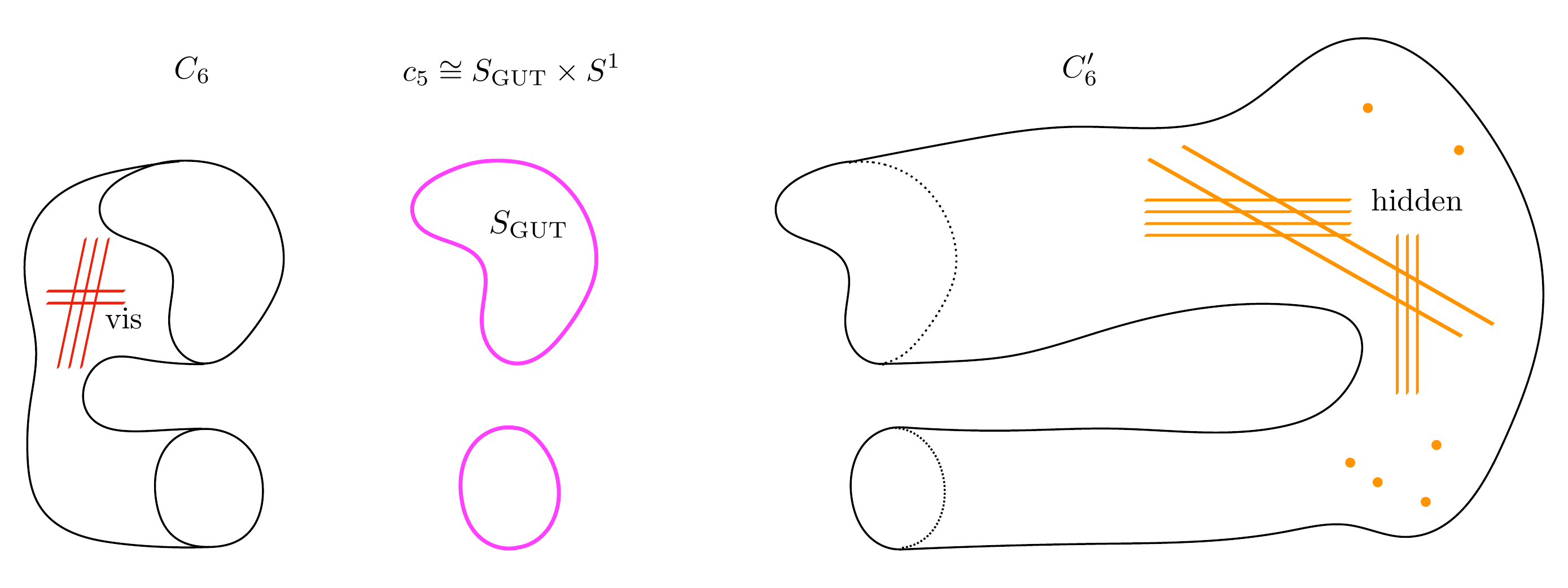}
\end{center}
\caption{
Depiction of different six-chains $C_6$ and $C_6'$ of the internal space bounded by the same five-cycle $c_5$ (pink) wrapped by our metastable five-branes. By suitably localizing the branes in the internal geometry, the volume of these
six-chains can be different, thus leading to a preferential decay mode to Standard
Model visible states (left) rather than hidden sector states (right), which can come from 7-branes (lines) or 3-branes (points).}
\label{fig:multisixchain}
\end{figure}

To see how this comes about, we return to our description of the
$\mathrm{Spin}(7)$ Universe in terms of a collection of $N$ discretized pixels
spanning our $S^3$. Recall that each such pixel originates from a five-brane
wrapped over a local five-cycle of the geometry. Here it is important to note
that such five-cycles may be present in the local geometry experienced by the
five-brane, but may be globally trivial. This is precisely the setup of
section 6.1 of reference \cite{Heckman:2018mxl}, where we take a stable
degeneration limit of a Calabi--Yau fourfold and in the tubular gluing region
identify a local five-cycle given by an $S^1 \times S_{\text{GUT}}$, with
$S_{\text{GUT}}$ a K\"ahler surface; see figure \ref{fig:multisixchain} for a
depiction of the internal geometry. Mathematically, the five-cycle bounds a
six-chain (a six-cycle with boundary), and the unwinding of the five-brane in
the geometry involves this process. Similar uses for relative homology cycles
have previously appeared in the string phenomenology literature, e.g.
\cite{Buican:2006sn, Verlinde:2006bc, Beasley:2008kw, Donagi:2008kj}.

\begin{figure}[t!]
\begin{center}
\includegraphics[trim={0cm 2.5cm 0cm 3cm},clip,scale=0.5]{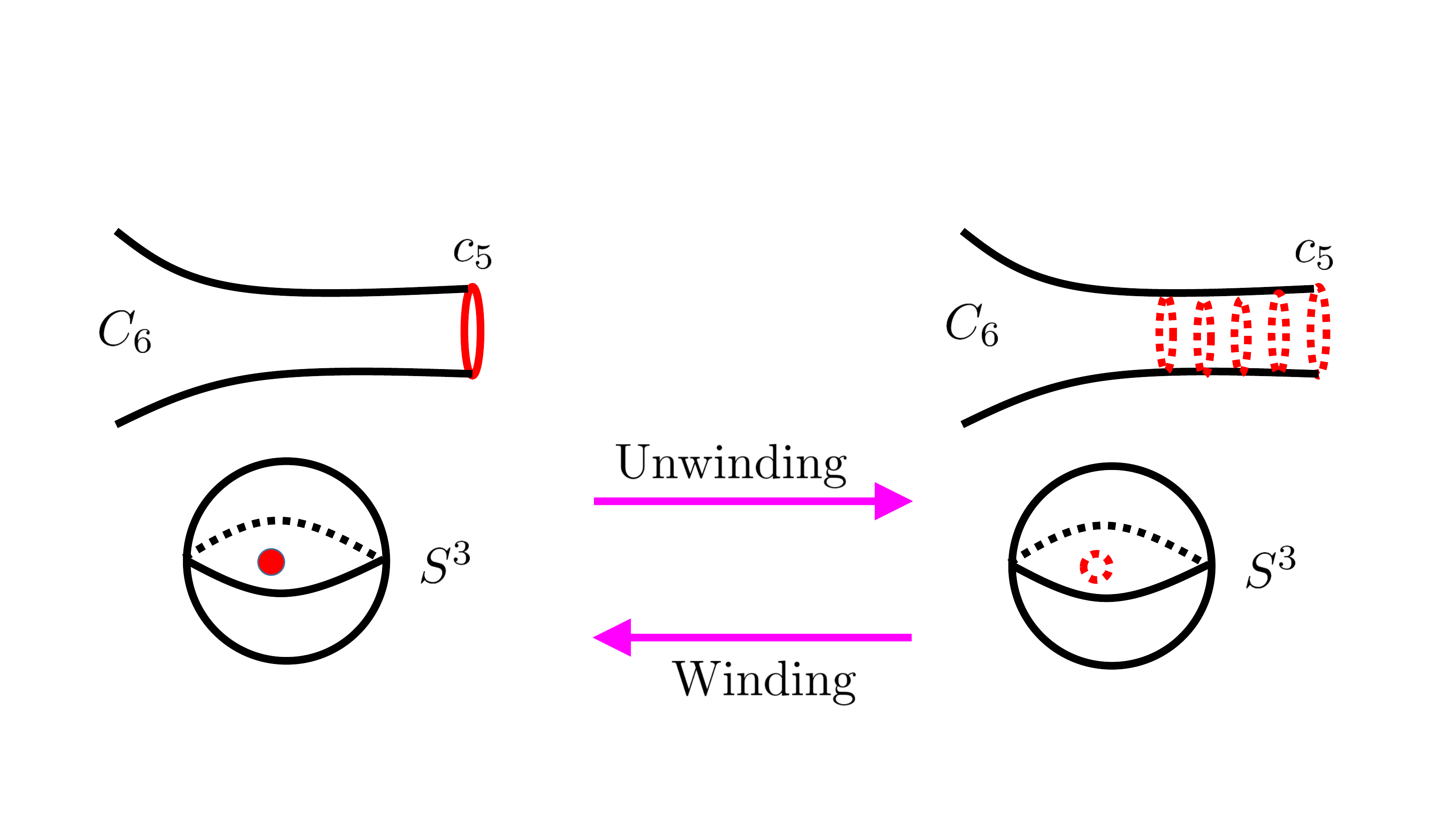}
\end{center}
\caption{
	Pixel winding and unwinding in the compactification geometry.
A local five-cycle $c_5$ which bounds a six-chain $C_6$ provides a decay mechanism for
five-branes in the geometry. Each five-brane defines a pixel in the 4D spacetime,
and its subsequent unwinding leads to a drop in the number of flux quanta.
Conversely, the number of pixels can increase through a winding process. The particular process
which dominates depends on whether the 4D spacetime is expanding (winding) or contracting (unwinding).
}
\label{fig:unwinding}
\end{figure}

In the global geometry there is nothing to prevent the five-branes from
unwinding since there is no globally conserved ``pixel charge''.  The value of
$N$ may therefore jump up or down after such a transition. Let us also note
that this is really a genuine tunneling event in the 1D system defined by
$\mathbb{R}_{\text{time}}$ and is not a bubble nucleation process as occurs in
a false vacuum (for example, the energy density before and after a jump are
the same). The tunneling process we are considering occurs even for vacua with
the same energy densities and does not involve a 4D field configuration. It is
more akin to an instanton process in Yang--Mills theory which can change the
total winding number ``at infinity.'' See figure \ref{fig:unwinding} for a
depiction of this winding and unwinding process in the geometry.

It is also instructive to consider the decay products of this unwinding.
Observe that a five-brane locally couples to a six-form potential $B_6$
via the worldvolume coupling
\begin{equation}\label{B6coupling}
  {\cal L}_{\rm five-brane} \supset \underset{\textrm{five-brane}}{\int} B_6 \,.
\end{equation}
In the 4D spacetime, the $B_6$ quanta are associated with the fluctuations of
the axion $\varphi$ introduced below equation (\ref{axionflux}). Consequently,
there is a decay process which converts a pixel into off-shell axions. These
will subsequently decay to other states from the Standard
Model and extra sectors. For example, in the 4D Lagrangian
there is a coupling to gauge fields with field strength $F_{\mu \nu}$
of the schematic form:
\begin{equation}
{\cal L}_{\rm 4D} \supset \int_{\rm 4D} \dd^{4}x \sqrt{-g} \, \frac{\varphi}{f_{\text{ax}}}
\mathrm{Tr} F^{\mu \nu} \widetilde{F}_{\mu \nu} \,.
\end{equation}
The details of this decay process clearly depend on how this axion couples to
various sectors, but a priori there does not seem to be any issue with it
preferentially decaying to visible sector states.  For example, if the extra
sector gauge fields are all strongly coupled at high scales (as is generically
the case) the scale of confinement will be quite high, and this gaps out those
excitations from the low energy spectrum.

``Winding up'' additional five-branes can also occur. Recall from equation (\ref{naboo})
that the internal profile of the dilaton sets the value of the cosmological constant. A
local change in this gradient amounts to ``pumping'' five-branes into the system, i.e. it is
the condensation of gravitational potential energy stored in the extra dimensions into five-branes.
From a 4D perspective, this conversion process can be summarized by saying that the ground state contains
hole and pixel excitations, and this process captures the conversion of a hole to a pixel.
Again, we expect this to be accompanied by relativistic radiation due to the coupling of line
(\ref{B6coupling}). See figure \ref{fig:twolevelnew} for a depiction of the
decay products from pixel winding and unwinding.

We also remark that there could be multiple six-chains which are all bounded
by the same five-cycle. Some of these six-chains may overlap in the geometry
where the Standard Model is localized, while others will be far away from the
Standard Model.  By tuning the metric of the internal geometry in these
patches, it should also be clear that the decay rate to the Standard Model
relative to other extra sectors can also be enhanced.  See figure
\ref{fig:multisixchain} for a depiction of different six-chains which can
bound a given metastable five-cycle.

Let us now turn to an estimate for the rate of these winding and unwinding
processes.  Topologically, we view the five-brane as bounding a six-chain
$C_6$.  The transition rate for a single pixel to come into or out of
existence is then controlled by
\begin{equation}
  \Gamma_{X} \sim M_{\ast} \exp\left[-M_{\ast}^6 \, \mathrm{Vol}(C_6)\right] \,,
\end{equation}
where $M_{\ast}$ is a characteristic string scale mass which can depend on
details of the model, including for example the tension of the five-brane, and
the volume of the local five-cycle it wraps. In principle, $M_{\ast}$ can
differ from the string scale $\ell_{s}^{-1}$ by a power of the string
coupling, but in the context of F-theory this is typically an order one
number. For this reason we shall typically neglect the distinction between
$M_{\ast}$ and $\ell_{s}^{-1}$.  With this in mind, the control parameter for
the decay rate is the volume of the six-chain. Observe that in the limit where
we decompactify the internal directions, this becomes infinitely large and the
transition rate vanishes. Note also that even a moderate increase in the
volume of the six-chain relative to the string scale will lead to a
significant suppression in $\Gamma_{X}$. For example, a well-motivated value
for the volume of the six-chain $C_6$ would be
\begin{equation}
  \mathrm{Vol} (C_6) \sim \frac{1}{M_{\text{GUT}}^{4} M_{\ast} M_{\bot}} \,,
\end{equation}
where $M_{\bot}^{-1}$ is the length of the direction normal to the five-cycle.
The appearance of the GUT scale is just one well-motivated possibility, and
originates from the assumption that the five-cycle takes the form $S^{1}
\times S_{\text{GUT}}$ with $S_{\text{GUT}}$ the four-cycle used to engineer a
GUT model. Clearly, the exponent of the decay rate can easily range from
$10^{2}$ to much larger values, depending on the values of these volumes.

We have already remarked that at least at the ``top of the hill'' we expect
the internal volume to be a tunable parameter relative to the size of the 4D
spacetime $S^3$. There will, of course, still be some implicit dependence on
$N$ and it is important to understand what such winding/unwinding events will
do to the volume of the internal geometry, especially as it might impact the
4D value of Newton's constant. Observe, however, that changing the number of
five-branes in the system is accompanied by a release in radiation. This
conservation of energy means that during all of these transitions, the 4D
Newton constant will remain the same. Said differently, once the proper
relation between the 4D Newton constant and volume of the internal
six-manifold $\mathcal{B}$ (the base of the F-theory model) has been set at
the top of the hill,
\begin{equation}
  \frac{1}{G_{\text{Newton}}} \sim M_{\ast}^8 \mathrm{Vol}(\mathcal{B}) \,,
\end{equation}
it remains unchanged even after winding/unwinding events.

\subsection{Two Level System}

It is also instructive to model the generic winding and unwinding processes in
terms of a two level quantum mechanical system, namely a pixel state
$|\text{pix}\rangle$ and a hole state $| \text{hol} \rangle$, i.e. the absence
of a pixel. This ignores possible excited pixel and hole states which we
expect to play an important r\^{o}le during reheating (see section
\ref{sec:REHEATING}).  We view the Universe as having access to an infinite
reservoir of particles and holes, in line with the fact that pixel number is
not conserved. The Hamiltonian is captured by a $2\times2$ matrix%
\begin{equation}
\widehat{\mathcal{H}}_{\text{two-level}}=\left[
\begin{array}
[c]{cc}%
E_{\text{hol}} & \Gamma_{X}\\
\Gamma_{X}^{\ast} & E_{\text{pix}}%
\end{array}
\right]  \,.\label{twoham}%
\end{equation}
By abuse of notation, we include operators involving radiation in our
definition of $\Gamma_{X}$.

The energy splitting $E_{\text{hol}}-E_{\text{pix}}$ between the two states
depends on whether we are on the collapsing or expanding branch. In order to
determine it, we return to our expression for the energy density of the
pixels:
\begin{equation}
\rho_{\text{pixel}}(N)=\rho^{0}_{\text{stiff}}(1)\frac{N^{2}}{a^{6}}+\rho
^{0}_{\Lambda}(1)\frac{1}{N} + \rho_\text{curvature}^0(1)\frac{1}{a^{2}} \,,
\end{equation}
where here, the energy density $\rho_{\text{pixel}}(N)$ depends on two
parameters: the number of flux units $N$, as well as the scale factor $a$.
Note that the quantities $a$ and $N$ are only related to one another
``at the top of the hill.''

The energy density splitting between particles and holes is then given by the
difference%
\begin{equation}
\rho_{\text{split}}(N)=\rho_{\text{pixel}}(N+1)-\rho_{\text{pixel}}%
(N)=\rho^{0}_{\text{stiff}}(1)\frac{2N+1}{a^{6}}-\rho^{0}_{\Lambda}(1)\frac{1}%
{N(N+1)} \,.\label{splitlevel}%
\end{equation}
For a flux jump to occur, we require that the energy density stored
in the pixels after the jump is strictly smaller than before, the remaining
balance being provided by the produced radiation. Observe, however, that the
level splitting
\begin{equation}
	\rho_\text{split} = \rho_\text{pixel}(\text{after}) - \rho_\text{pixel}(\text{before}) \,,
\end{equation}
is strictly positive at the ``top of the hill.'' This is because
$\rho_\text{pixel}(\text{before})$ vanishes, and
$\rho_\text{pixel}(\text{after})$ is still a positive number (at
least for small flux jumps). This in turn means that no jump can occur, either
to more or less flux quanta since it is entropically forbidden.

\begin{figure}[t!]
\begin{center}
\includegraphics[trim={0cm 0.5cm 0cm 0.5cm},clip,scale=0.5]{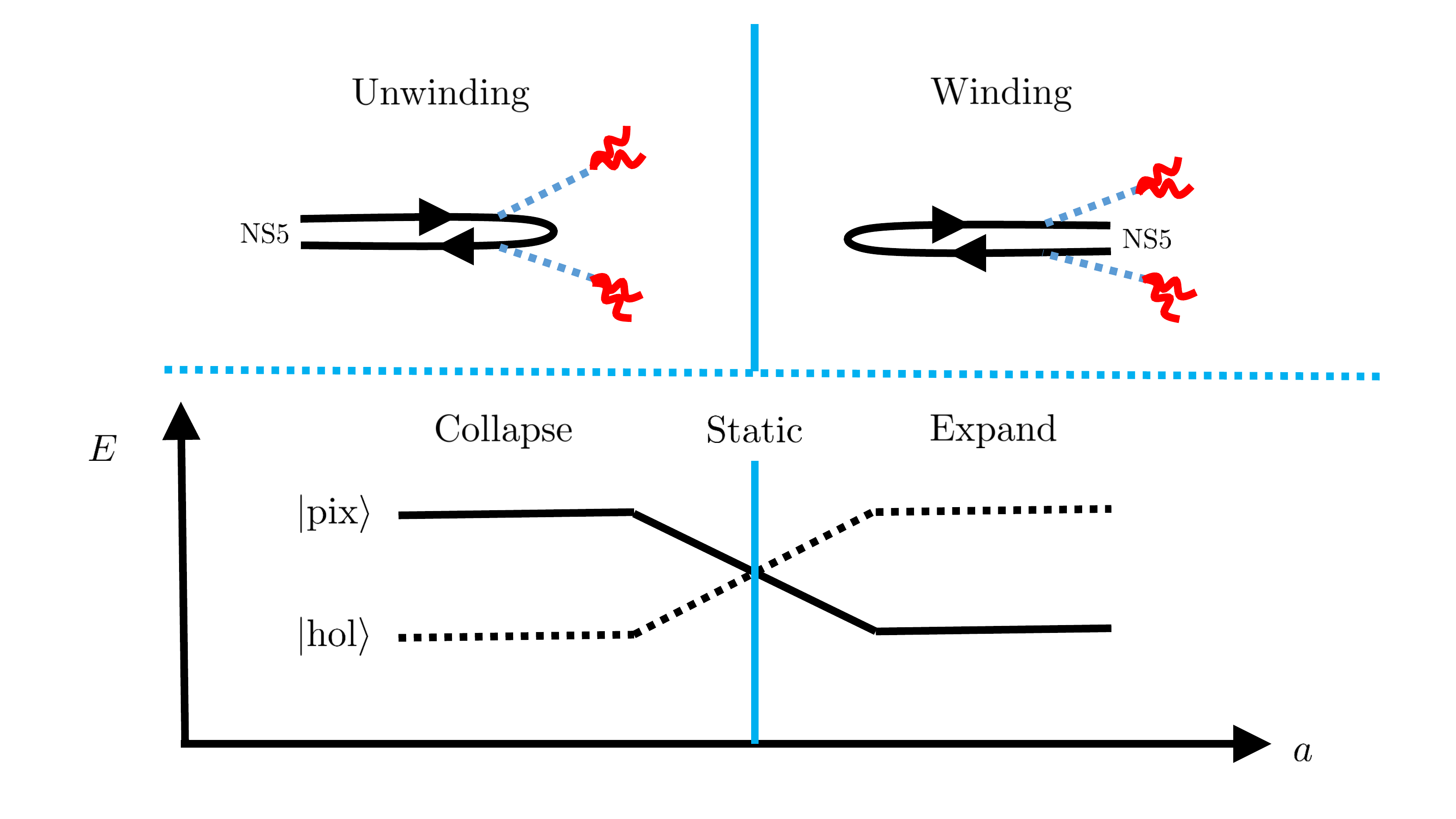}
\end{center}
\caption{
	Depiction of the two level system for pixels and holes. On the collapsing branch, the pixels have higher energy while on the expanding branch the holes have higher energy. In the stringy description, unwinding a five-brane corresponds to the creation of a hole while winding up a five-brane corresponds to the creation of a pixel. The blue dashed lines in the upper panel represent off-shell axions which subsequently decay to radiation. Energy level splittings are not drawn to scale in the figure.
}
\label{fig:twolevelnew}
\end{figure}

This also means that on the collapsing branch $E_{\text{pix}%
}>E_{\text{hol}}$ while on the expanding branch $E_{\text{hol}} >
E_{\text{pix}}$. An excited state can transition to a lower state by
  releasing radiation.  This is the transition rate process we already
  mentioned. See figure \ref{fig:twolevelnew} for a depiction of this two
  level system.

\section{Flux Jumps} \label{sec:JUMP}

Having seen that the lifetime of our holes/pixels can be engineered to be
long-lived, we now ask about possible jumps in the number of flux units. Since
we anticipate having an aggregate of such pixels, it is more appropriate to
view $N$ as a function of time, $N(t)$. What we would like to know is whether
we can expect to gain or lose pixels as the Universe evolves.

As a first step in addressing possible jumps in the number of flux units, we
consider a semi-classical analysis in which all pixel transitions happen
sufficiently slowly relative to the value of the Hubble parameter $H$. In this
case, we can track the effects of pixel creation and annihilation by computing
the effective energy densities in the system. We view the expansion of the
Universe as a classical process and therefore assume that the scale factor $a$
as well as its first and second derivatives are continuous functions of
time.\footnote{If we attempt to treat the decay as instantaneous the second
derivative will be discontinuous.} We do not, however, assume that higher
order derivatives are smooth. In fact, we can also extend our analysis to the
case of discontinuous jumps (see Appendix \ref{app:DISCO}).

On general grounds, we expect to execute a jump in the flux units, and that
there is also an accompanying change in the amount of radiation. On a given
branch, the energy density for the stiff fluid and cosmological constant
depend on the number of flux quanta $N$:
\begin{equation}
\rho_{\Lambda}(N) = \rho^{0}_{\Lambda}(1) \frac{1}{N}\, \,\,\, \text{and}
\,\,\, \rho_{\text{stiff}}(N) = \rho^{0}_{\text{stiff}}(1) \frac{N^2 }{a^6} \,.
\end{equation}
As in our discussion near equation (\ref{splitlevel}), we have explicitly
indicated the $N$ dependence of the various energy densities but have left
implicit the scale factor dependence. Since each burst of radiation increases
the entropy, we see that the creation of radiation is entropically favored.

Suppose now that our Universe has some component of stiff fluid, cosmological
constant, and radiation (as well as curvature) with flux number $N$ and
proceeds to transition to a Universe with flux number $M$. Prior to the jump,
the Hubble parameter is given by
\begin{equation}\label{beforejump}
H^2 = - \frac{1}{a^2} + \frac{8 \pi G}{3} \left( \rho^{0}_{\text{stiff}}(1)
\frac{N^2 }{a^6} + \rho^{0}_{\Lambda}(1) \frac{1}{N} +
\frac{\rho^{0}_{\text{rad}}(N)}{a^4} \right) \,.
\end{equation}
Just after the transition $N \rightarrow M$, the scale factor
and Hubble parameter are identical so we also have
\begin{equation}\label{afterjump}
H^2 = - \frac{1}{a^2} + \frac{8 \pi G}{3} \left( \rho^{0}_{\text{stiff}}(1)
\frac{M^2}{a^6} + \rho^{0}_{\Lambda}(1) \frac{1}{M} +
\frac{\rho^{0}_{\text{rad}}(M)}{a^4} \right) \,.
\end{equation}

Since the scale factor and Hubble parameter are assumed to be
the same during this transition, we can subtract the two quantities
to obtain the difference in radiation:
\begin{equation}\label{radiationdifference}
\Delta \rho_{\text{rad}}
= \rho^{0}_{\Lambda}(1) \left( \frac{1}{N} - \frac{1}{M} \right)
+ \rho^{0}_{\text{stiff}}(1) \left(\frac{N^2}{a^6} - \frac{M^2}{a^6} \right)
\,.
\end{equation}
It is convenient to write this equation as
\begin{equation}\label{diffDelta}
\Delta \rho_{\text{rad}} = - \Delta \rho_{\Lambda} - \Delta
\rho_{\text{stiff}} \,,
\end{equation}
in the obvious notation (and switching to $a$-dependent densities). Note that
for $N \neq M$, the righthand side always has one positive term and one
negative term. From this, we can already see that radiation is either produced or destroyed
as a result of a tunneling event. If the number of flux units changes by a
small amount, then the radiation again dissipates rapidly, and we expect to
remain on a branch dominated by a cosmological constant (expanding branch) or
a stiff fluid (collapsing branch).  In the case of a large flux jump, however,
it is possible to produce a radiation dominated phase. Our main interest here will be on flux jumps that produce radiation, so we analyze the case with
\begin{equation}
  \Delta \rho_{\text{rad}} > 0 \,.
\end{equation}
The reverse process is entropically disfavored.

To proceed further, it is helpful to split up our analysis into different
cases for transitions that generate radiation, depending on whether $M > N$
or $N < M$. We shall first assume that all transitions take place in a regime
where the Universe is still macroscopically large, as is appropriate on the
expanding branch, as well as early on the collapsing branch. We return to an
analysis of the collapsing branch in section \ref{sec:COLLAPSO} and study
reheating after a large flux jump in section \ref{sec:REHEATING}.

\subsection{Jumping Up \label{ssec:jumpup}}

Returning to equations (\ref{radiationdifference}) and (\ref{diffDelta}), we
find that when $M> N$, $\Delta \rho_{\text{stiff}} > 0$ while $\Delta
\rho_{\Lambda} <0$. To generate radiation from this transition, the condition
$\Delta \rho_{\text{rad}} > 0$ means
\begin{equation}
  - \Delta \rho_{\Lambda} > \Delta \rho_{\text{stiff}} \,,
\end{equation}
or
\begin{equation}
\frac{\rho^0_{\Lambda}(1)}{\rho^0_{\text{stiff}}(1)} > \frac{1}{a^6} M N (M+N)
\,.
\end{equation}
This occurs automatically when $a$ is sufficiently large, i.e. we are far
along on the expanding branch.  It is also not possible to satisfy this
sufficiently far on the collapsing branch. Indeed, this just follows from the
location of the ``top of the hill'' $\overline{a} \sim \sqrt{N} \ls$.  Of
course, we ought not to trust our analysis if we are too far down on the
collapsing branch since in this case there are higher-derivative corrections
to the Einstein--Hilbert action. We conclude, then, that on the expanding branch, jumping up to a higher value
of flux quanta is substantially preferred, though it is also possible even in
the early time collapsing branch.

We can also see that a rare event can occur which generates a large transition
in radiation. For radiation to dominate over the energy density from the
cosmological constant on the expanding branch, we also require
\begin{equation}
  \Delta \rho_{\text{rad}} > \rho_{\Lambda}(M) \,,
\end{equation}
which means
\begin{equation}
  0 > \Delta \rho_{\text{stiff}} > \Delta \rho_{\Lambda} + \rho_{\Lambda}(M) \,.
\end{equation}
So a necessary condition for a radiation dominated jump to occur is that
\begin{equation}
  M > 2 N \,,
\end{equation}
so the size of the jump $M - N$ must be at least of order $N$. This is already
a rare event, so we expect $M \sim N$ as an order of magnitude estimate.

\subsection{Jumping Down}

Consider next the case of a downward jump in the number of flux units. As we
have already mentioned, a semi-classical treatment is only valid provided the
scale factor is sufficiently large. Modulo this caveat, the analysis of
inequalities is much as in the case of an up jump. A generic jump
down will occur on the collapsing branch, and simply leads to a phase which is
dominated even more by the stiff fluid. On the other hand, we can also see
that there are definite limitations since as the scale factor collapses to
zero size, we must include the quantum fluctuations associated with this
operator as well. We now turn to a brief discussion of this system.

\section{The Collapsing Branch} \label{sec:COLLAPSO}

In the previous section, we saw that in some cases, starting at the ``top of
the hill'' of the static solution can roll on to the collapsing branch. Here we
study some aspects of this collapsing branch.  Our discussion is necessarily
more cursory because it requires us to discuss some additional details on the
quantum structure of the scale factor. Even so, we shall aim to isolate some
aspects that are robust. Along these lines, we first argue that
tunneling between different values of flux quanta persists in this regime.
Second, we show that under suitable assumptions, the scale factor is expected
to fluctuate back out to large values.

Deep on the collapsing branch, it is more appropriate to work in terms of a
dimensionally reduced 1D action.  With this in mind, our starting point is the
4D action
\begin{equation}
S_\text{grav, 4D}=\frac{1}{16\pi G_{N}}\int \dd^{4}x\,\sqrt{-g}\,\left[  R-\frac{1}%
{2}|H_{3}|^{2}-2\Lambda\right]  \,.
\end{equation}
Introduce the lapse function $\mathcal{N}(t)$ via:
\begin{equation}
\dd s^{2}=-\mathcal N(t)^2\,\dd t^{2}+a(t)^{2}\dd \Omega_{S^3}^{2}\,,
\end{equation}
The resulting 1D action for the scale factor in the ``mini-superspace approximation''
is \cite{Heckman:2018mxl}:
\begin{equation}
S_\text{1D}=\frac{3\pi}{4G_{N}}\int \dd t\, a^3\left[  \mathcal N\frac{1}{a^2}-\frac{\mathcal N}{a^{6}%
}\frac{(\dot \varphi - N )^{2}\ls^{4}}{48\pi^{4}}-\mathcal N^{-1}\frac{\dot{a}^{2}}{a^2}-\frac{1}{3}%
\mathcal N\,\Lambda\right] + ... \,,\,
\end{equation}
where here, we have added a ``...'' to capture the fact that there are many
higher-derivative terms which scale inversely in the scale factor $1/a$, as
well as higher-order derivative and potential terms for the axion.

Though we ought to treat the dynamics of the axion and scale factor
simultaneously, we can already see that there is no obstruction to flux
tunneling in the small $S^3$ limit.  To track this, introduce a number
operator $\widehat{N}$ defined by the total number of pixels, i.e. the
integral of our flux over the three-sphere:
\begin{equation}
  \widehat{N} = \frac{1}{ \ls^2} \int_{S^3} H_3 \,.
\end{equation}
Introducing the conjugate momentum operator $\widehat{P}_{\varphi}$ to the
axion, we see that the Hamiltonian for this 1D system includes the terms
\begin{equation}
\widehat{\mathcal{H}}_{\text{ax}} = \frac{1}{2} (\widehat{P}_{\varphi} - \widehat{N})^2 + V(\widehat{\varphi})
+ \cdots \,,
\end{equation}
where the ``$\cdots$' refers to all dynamics including the scale factor and
the pixels. The potential energy $V(\widehat{\varphi})$ is generated by
non-perturbative effects involving the coupling of the axion to various gauge
theory sectors (as well as gravity).  We take this to mean that we can label
vacua by $\vert N \rangle$, and that there are transitions between different
values of $N$. At a semi-classical level, we also expect that as we move
deeper on the collapsing branch that $N$ may unwind to an order one value. In
this case, the overall size of the $S^3$ will also be string scale.  This also
means a jump in the value of the scale factor can easily push us back out onto
the expanding branch.

To support this qualitative assertion consider next the quantum dynamics of
the scale factor.  It is helpful to work in terms of a canonically normalized
kinetic term for the fluctuations of the scale factor. Introducing a field
$A(t)$ such that
\begin{equation}
  a^{3/2} = \frac{3}{2 \sqrt{2}} A \equiv \alpha A \,,
\end{equation}
the 1D action takes the form
\begin{equation}
S_\text{1D}=\frac{3\pi}{4G_{N}}\int \dd t\left[  \alpha^{2/3} A^{2/3} \mathcal N-\frac{\mathcal N}{\alpha^2 A^2
}\frac{(\dot \varphi - N )^{2}\ls^{4}}{48\pi^{4}}-\mathcal N^{-1}\frac{1}{2}\dot{A}^{2}-\frac{1}{3}%
\mathcal N\,\alpha^2 A^{2}\,\Lambda\right] + ... \,,\,
\end{equation}
There are a number of conceptual subtleties involved in the quantum mechanical
interpretation of this system.  For one, we see that the field $A(t)$ has a
wrong sign kinetic term so it is not entirely clear whether one should expect
a well behaved ground state for the system. For some discussion on the
different interpretations from a bottom up point of view, see e.g.
\cite{DiazDorronsoro:2017hti, Feldbrugge:2018gin}. From a top down point of
view there appears to be no issue with interpreting this field in the
accompanying stringy worldsheet theory \cite{Heckman:2018mxl}.

At a qualitative level, then, we expect the quantum dynamics for the scale
factor to be controlled by a Hamiltonian operator of the schematic form
\begin{equation}
  \widehat{\mathcal{H}}_{\text{scl}} = - \frac{1}{2}P^2_{\widehat{A}} + V(\widehat{A}) \,.
\end{equation}
The variation of the lapse function enforces the constraint that we only study
states which are annihilated by this operator:
\begin{equation}
  \widehat{\mathcal{H}}_{\text{scl}} \vert \Psi \rangle = 0 \,.
\end{equation}
As a passing comment, we observe that if we only include the contribution from
the stiff fluid, we get a conformal quantum mechanics system with a ``wrong
sign kinetic term.''

The main qualitative point we wish to emphasize is that there is no sense in
which the wave function is completely concentrated at $A = 0$. Indeed, in the
simple case $V(A) \sim 1/A^2$ we find a wave function which has substantial
support at large values of $A$, and actually must be cut off at large values
to ensure it is normalizable. We take this to mean that even if the system is
``unlucky'' and initially rolls to a collapsing phase, it can eventually
bounce back out to a large scale factor.

Note also that there is no issue with the axion winding around several times,
causing a corresponding change in the flux quanta along each tunneling
transition. Again, we take this to mean that even in the ``worst case
scenario'' where $N$ becomes an order one number, it can eventually transition
back to larger values. In fact, we see that if we pass onto a collapsing
branch and first unwind to small $N$, then a subsequent transition in the
value of the scale factor makes it easier to instead jump back out to the
expanding branch of the solution.

Summarizing, we see that it does not really matter whether the $\mathcal{N} =
1/2$ supersymmetric solution initially collapses or expands. Eventually, it
will find its way out to the expanding branch anyway. This also means the
scenario is relatively insensitive to initial conditions, though it would of
course be desirable to determine the probability distribution in $N$.

\section{Reheating} \label{sec:REHEATING}

From the discussion of the previous section, we see that the Universe will
inevitably wind up on the expanding branch. Typically, we expect a slow
increase in the number of pixels, but occasionally a rare event will occur
where a large number of holes transition to pixels, releasing a large amount of radiation.
For ease of exposition, we refer to this radiation as \textquotedblleft
photons\textquotedblright\ though really any relativistic matter could appear
here, including gravitational waves. In this regime, we expect the dynamics of
pixel production to be qualitatively different.

When this occurs, we expect out of equilibrium dynamics to trigger an
exponential increase in the number of pixels. Since we have only imperfect
knowledge of the string compactification as well as the pixel dynamics, our
discussion will necessarily involve some speculative aspects.  We leave a more
detailed analysis of some elements for future work.

With these stated caveats, let us now turn to the production of pixels in the
regime with a large amount of background radiation.  The energy density stored
in the released radiation is controlled by the size of the jump, which, as we
have already mentioned, is the smallest value of $N$ which permits a
transition to radiation domination (otherwise we are, by definition, still in
the cosmological constant dominated phase). The energy density stored in
radiation at the end of inflation is then on the order of%
\begin{equation}
\rho_{\text{rad}}\sim\frac{\rho_{\Lambda}^{0}(1)}{N_{\text{ignite}}%
} \,.\label{Nignite}%
\end{equation}
Assuming a roughly thermal spectrum, we also know that the energy density and
number density for photons scales with temperature as
\begin{equation}\label{raddude}
  \rho_{\text{rad}}\sim T^{4} \quad \text{and} \quad n_{\text{rad}}\sim T^{3} \,,
\end{equation}
where here, we have dropped all ``order one factors,'' including $g_{\ast}$
and $g_{\ast S}$. Our reason for doing so is that we have imperfect knowledge
of the thermodynamics at this early epoch. Returning to equation
(\ref{Nignite}), we can extract the initial temperature at the start of
ignition,
\begin{equation}
T_{\text{ignite}}\sim\left(\frac{\rho_{\Lambda}^{0}(1)}{N_{\text{ignite}}}\right)
^{1/4} \,,
\end{equation}
and the number density therefore scales with temperature and $N_{\text{ignite}}$ as
\begin{equation}
n_{\text{rad}}\sim \left(\frac{\rho_{\Lambda}^{0}(1)}{N_{\text{ignite}}}\right)
^{3/4}\times\left(  \frac{T}{T_{\text{ignite}}}\right)  ^{3}%
\,.\label{numberdensity}%
\end{equation}

What we need to understand is the impact of this sudden increase in radiation
on the population of holes and pixels. Clearly, we are not in an equilibrium
configuration for the holes and pixels, so we can expect the production rate
of pixels to speed up during this process. Below, we sketch how the rapid
increase in radiation facilitates a corresponding increase in the production
of pixels.

\begin{figure}[t!]
\begin{center}
\includegraphics[trim={0cm 1.5cm 0cm 1.5cm},clip,scale=0.5]{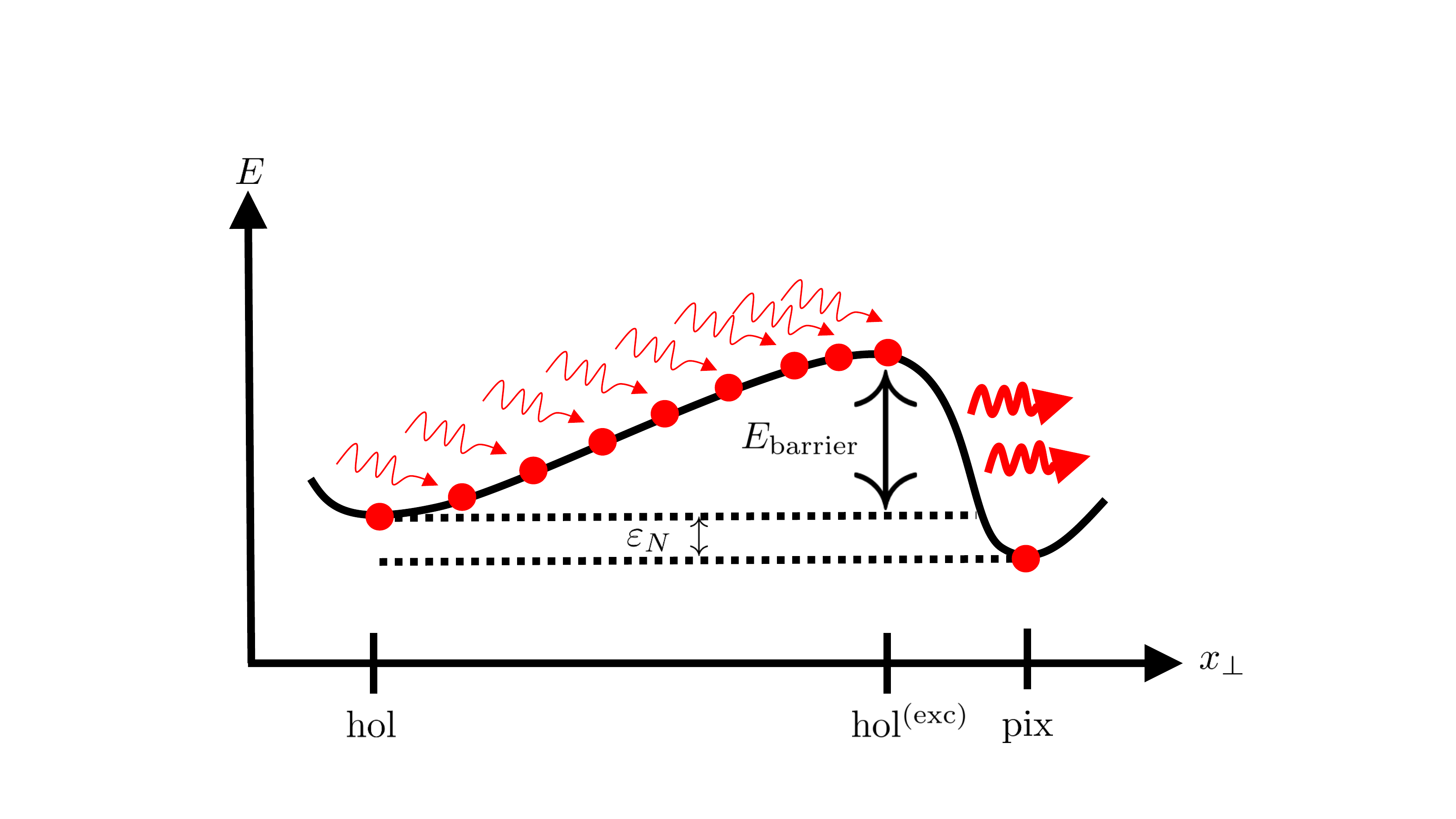}
\end{center}
\caption{Depiction of the energy barrier for a hole winding up to a pixel in the presence of background radiation. The x-axis $x_\bot$
denotes the direction orthogonal to the five-brane. Soft photons hitting a hole push it up the energy barrier until it eventually
reaches an excited hole state which can quickly decay to a pixel. This final decay releases radiation. A consequence of this
process is that low energy photons are converted to high energy photons, leading to an increase
in the effective temperature in the system. This in turn increases the production of pixels.}
\label{fig:freebarrier}
\end{figure}

To study this in more detail, we first display in figure \ref{fig:freebarrier}
a sketch of the expected energy landscape to convert holes to pixels. As is
evident from the figure, even though there is a rather small level splitting
between holes and pixels, there is a large energy barrier which must first be
overcome to transition between the two states. This energy barrier reflects
the large barrier in winding/unwinding branes. At a qualitative level, we
expect this barrier to be asymmetric because of the detailed geometry of the
internal directions.  Let us also note that even though we expect $G_{\text{Newton}}$
to be unaffected by these transitions, the local profile of the internal geometry will surely be
affected by the change in the value of $N$. This also means we should expect
a transition rate exponent which depends on $N$.

The difference in energy densities between pixels and holes was already
estimated in equation (\ref{splitlevel}), and on the expanding branch it is
given by
\begin{equation}
  \rho_{\text{split}}(N)\sim\rho^{0}_{\Lambda}(1)\frac{1}{N(N+1)} \,,
\end{equation}
which is a very small number. Since there are $N$ pixels, we conclude that the
energy splitting scales with $N$ as%
\begin{equation}
  E_{\text{split}}(N)\sim\frac{M_{\text{pl}}}{N^{3}} \,,
\end{equation}
a very small number indeed. On the other hand, the height of the barrier
depends on the volume of the five-cycle wrapped by a five-brane, and the
difference between the biggest and smallest volume five-cycle:
\begin{equation}
E_{\text{barrier}}\sim M_{\ast}^{6}\left[  \text{Vol}(M_{5}^{\text{big}%
})-\text{Vol}(M_{5}^{\text{small}})\right] \,.
\end{equation}
This can in principle be at the GUT scale $\sim 10^{16}$ GeV or far lower,
depending on details of the geometry. One should view this as a
tunable parameter.

Now, in the absence of background radiation, we have already explained that
there is a decay process in which a hole winds up to form a pixel, i.e.
standard barrier penetration in quantum mechanics. Thermal effects make it
easier to overcome this barrier, but due to the asymmetric shape of the
barrier, namely, the presence of a chemical potential, we expect a
preferential increase in pixel final states rather than the reverse process.

Observe that each hole will be bombarded by a number of soft photons, and each
collision takes the hole to a slightly more excited state. If the number of
soft photons is sufficiently high, we expect a conversion rate from holes to
pixels. The final transition from an excited hole to a pixel (at the top of
the free energy barrier) is comparably fast, and leads to a release of energy
absorbed from the stored photons. Note also that the reverse process of pixel
conversion to holes is far less frequent. This is because a pixel would need
to quickly climb the potential barrier, and thus requires a more energetic
photon. What this means is that as the holes convert to pixels, the spectrum
of photons is tilted up to more energetic photons. As this occurs, the rate of
hole to pixel conversion also increases.

This is an out of equilibrium process, but eventually the spectrum of photons
becomes so energetic that a balance is reached between the production and
destruction of pixels. This occurs when the temperature becomes comparable to
the barrier potential:
\begin{equation}
  T_{\text{reheat}}\sim E_{\text{barrier}} \,.
\end{equation}
At this point pixel production stops, and we enter into a standard freeze-out
scenario:\ The Universe expands and as it does, it cools down. Eventually the
holes/pixels freeze-out as no photons can hit the pixels. This occurs at a
freeze-out temperature comparable to the mass of the pixels (see e.g.
\cite{Kolb:1990}),
\begin{equation}
  T_{\text{freeze}}\sim 0.1 m_{h} \,,
\end{equation}
where $m_h$ is the mass of the hole state. Note that $m_{h}<E_{\text{barrier}}$.

\begin{figure}[t!]
\begin{center}
\includegraphics[trim={0cm 5cm 0cm 5cm},clip,scale=0.5]{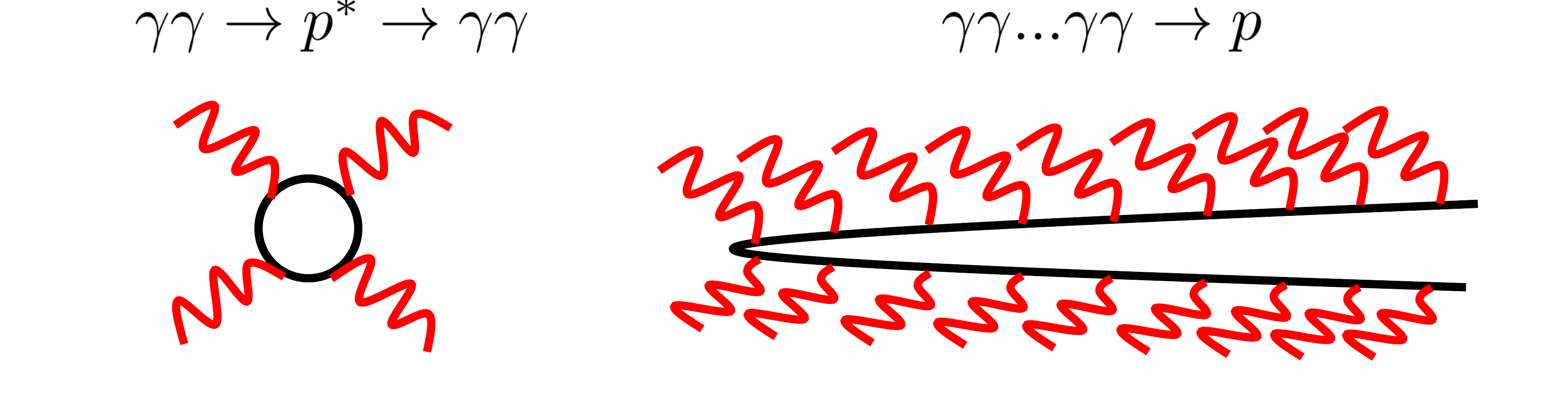}
\end{center}
\caption{
	Depiction of low energy background radiation scattering off of holes. This leads to the creation of a virtual
pixel (left). If, however, enough radiation is present, the hole excitation will pass the free energy barrier and
the pixel will persist (right).}
\label{fig:softserve}
\end{figure}

We would now like to estimate the production in pixels as the Universe passes
from the initial stages of ignition to reheating. To figure out the pixel
production rate, we first ask how a hole can convert to a pixel through photon
collisions. Typically, we expect to produce a virtual pixel, and for it to
subsequently decay back into a hole (see figure \ref{fig:softserve}).
Treating the hole as the ground state, the relevant process is%
\begin{equation}
  \gamma\gamma\rightarrow p^{\ast}\rightarrow\gamma\gamma \,.
\end{equation}
Treating the hole as an excitation in its own right, we can also view this as
hole/photon Compton scattering,
\begin{equation}
  \gamma h\rightarrow\gamma h^{\prime} \,.
\end{equation}
The cross section for this process clearly scales inversely in the mass
squared of the hole%
\begin{equation}
  \sigma_{\gamma h\rightarrow\gamma h^{\prime}}\sim\frac{1}{m_{h}^{2}} \,.
\end{equation}
Here, we have really presented the inclusive cross section as obtained by
integrating over the momentum distribution of initial and final state photons
and holes. These are important pre-factors, but given our limited
understanding of the physics of holes, we leave it in this general form.
Presumably this could be estimated by calculating the greybody factors for our
five-branes wrapped on metastable five-cycles.

To get a conversion to a pixel involves many such scattering events. With this
in mind, we can expect a non-trivial temperature dependence through the
dimensionless ratio $T/m_{h}$. On the other hand, since the temperature of the
photon radiation at equilibrium is expected to be roughly of order $m_{h}$
anyway, we shall simply assume that the process to convert a hole to a pixel
via photon scattering is of a similar order of magnitude \textquotedblleft up
to order one factors\textquotedblright:
\begin{equation}
  \sigma_{\gamma h\rightarrow\gamma p}\sim\frac{1}{m_{h}^{2}} \,.
\end{equation}
Here again, we have taken an inclusive cross section. We of course expect the
overall pre-factor for this process to differ from that of photon/hole
scattering to an excited hole state, but since we can only reliably extract
the overall mass scaling dependence, we have suppressed all of these
additional details.

Let us now turn to the rate of pixel production. As we have already mentioned,
the production of each pixel leads to a corresponding tilt in the energy
distribution of photons to a higher temperature. Since the number of holes is
in some sense \textquotedblleft infinite,\textquotedblright\ it is
conceptually simplest to work in terms of the number densities for just the
photons and pixels. With these remarks in place, we estimate the rate equation
for the number density of pixels and photons as%
\begin{equation}
\frac{\dd n_{p}}{\dd t}\sim\sigma_{\gamma h\rightarrow\gamma p}n_{\gamma}n_{p}\text{
\ \ and \ \ }\frac{\dd n_{\gamma}}{\dd t}\sim-\sigma_{\gamma h\rightarrow\gamma
p}n_{\gamma}n_{p} \,.
\end{equation}
which, for constant $n_{\gamma}$, would yield an exponential increase in the
pixel number and a corresponding depletion in the number density of photons.
Of course, there is also some implicit frequency (and thus time)\ dependence
for the photons: It is still expected to be a black body spectrum, but now in
the presence of a non-trivial chemical potential. Since we are dealing with an
out of equilibrium process, we again content ourselves with a rough order of
magnitude estimate for $n_{\gamma}$. Substituting in our expression from
equation (\ref{raddude}) with $T \sim T_{\text{ignite}}$, we get an estimate
for the breeding rate:
\begin{equation}
\Gamma_{\text{breed}}\sim\sigma_{\gamma h\rightarrow\gamma
p}n_{\gamma}\sim\frac{1}{m_{h}^{2}}\times
\left(\frac{\rho_{\Lambda}^{0}(1)}{N_{\text{ignite}}}\right)  ^{3/4} \,.
\end{equation}
There are many order one factors here which are raised to large exponents,
which makes obtaining a reliable estimate somewhat challenging.  Proceeding
anyway, however, we obtain the following very crude $N_{\text{ignite}}$
scaling for the breeding rate,
\begin{equation}
\Gamma_{\text{breed}}\sim M_{\text{pl}}\times\frac{\left(  M_{\text{pl}}%
^{2}/m_{h}^{2}\right)  }{N_{\text{ignite}}^{3/4}} \,.
\end{equation}
The process continues until the temperature increases so much that pixels and
holes convert back and forth, i.e. we reach a stage of equilibrium. So, we get
our final estimate on the number of pixels after reheating%
\begin{equation}
N_{\text{reheat}}\sim N_{\text{ignite}}\times\exp\left(  \Gamma_{\text{breed}%
}/T_{\text{reheat}}\right)  \,,
\end{equation}
where $T_{\text{reheat}}$ is on the order of the barrier potential
$E_{\text{barrier}}$.

To give some representative examples, we take $N_{\text{ignite}}%
\sim10^{12}$, 
$E_{\text{barrier}} \sim 10^{15}$ GeV in which case we get%
\begin{equation}
T_{\text{reheat}}\sim10^{15}\text{ GeV}, \quad \Gamma_{\text{breed}}\sim
10^{17}\text{ GeV} \,,%
\end{equation}
which would generate an exponential increase in the number of pixels after
first ignition. Note that this reheating temperature is near the upper bound
allowed by our crude estimate in line (\ref{Tignite}) under the assumption
that the Universe did not enter a contracting phase and thus that
$T_{\text{reheat}}$ is at most $T_{\text{ignite}}$.  Also note that such a
high breeding rate also strains the regime of validity for a 4D field theory
computation. Both the reheating temperature and the breeding rate depend on
additional adjustable parameters (e.g. $g_\ast$ and $g_{\ast\,s}$) but this at
least shows that it is plausible to get an exponential increase in pixels
relative to the value during inflation.

Clearly, there are many elements here which are of a qualitative nature.
However, as far as performing more precise calculations go, the main input
from stringy physics is the profile of the free energy barrier for hole to
pixel conversion. In principle, a purely 4D analysis of this phenomenon
should be possible.

\section{Pixel Matrix Model} \label{sec:PIXEL}

In this section we introduce a matrix model for the pixels of our spacetime.
This is of course in the general spirit of M(atrix) theory
\cite{Banks:1996vh}.  For some earlier discussions of de Sitter matrix models,
see e.g.  \cite{Banks:2006rx, Susskind:2011ap}. We view our analysis as
providing additional evidence for the proposal of reference
\cite{Heckman:2018mxl} that F-theory on a $\mathrm{Spin}(7)$ background can be
obtained from the backreacted limit of five-branes wrapped over metastable
five-cycles. Another outcome will be a (crude)\ match to some thermodynamic
properties of de Sitter space.

As already mentioned, our starting point is the theory of five-branes wrapped
over a metastable five-cycle. In reference \cite{Heckman:2018mxl} it was
convenient to view this as a collection of NS5-branes, but for our present
purposes it will be more straightforward to work in the S-dual frame in terms
of D5-branes. In either case, we expect the low energy limit to be governed by
a collection of heavy particles in the 4D\ spacetime which break half the
supersymmetry of the original 4D\ $\mathcal{N}=1$ theory. An important comment
is that if we had placed the five-branes in the vicinity of seven-branes, we
would have similar Neumann--Dirichlet boundary conditions to that of D0-branes
in the presence of D8-branes and D6-branes. For supersymmetry to be preserved
in such a configuration requires the D0-branes to dissolve into flux, a point
we shall turn to shortly.

Now, although we must eventually allow the number of such particles to be
arbitrary, suppose first that we have $N$ such particles. We then expect to
have a 1D quantum mechanical system constructed
from $N\times N$ matrices in a $\mathrm{U}(N)$ gauge theory. There are four
directions in the 10D\ spacetime which are transverse to the five-brane so we
anticipate a matrix quantum mechanics theory with four adjoint valued scalars
$X^{i}$ with $i=1,...,4$. At the top of the hill, we anticipate that our
$\mathrm{Spin}(7)$ ansatz retains two real supercharges, so each $X^{i}$ is
the scalar component of a 2A multiplet, that is, each superpartner is a real
doublet of adjoint value fermions (see Appendix C of \cite{Heckman:2018mxl} as
well as the Appendix of \cite{Haupt:2008nu} for details). However, once the
spacetime starts rolling away from the top of the hill, supersymmetry will
inevitably be broken and the five-branes can wind and unwind.

This is the signal of a tachyonic instability, and related questions have been
addressed in the string theory literature in the context of the decay of
non-BPS\ D-branes \cite{Sen:1998ii, Sen:1998tt, Sen:1998ki, Sen:1999mg}.  At a
minimum, this means that we ought to include an additional adjoint valued
scalar $T$ which parameterizes the breaking of time translation invariance.
This operator is our tachyonic mode and should be viewed as a string which
stretches from the brane back to itself to execute
self-annihilation/reproduction. The semi-classical creation and destruction of
pixels was already treated in section \ref{sec:JUMP}. Since we will be
interested in the interplay between this tachyonic mode and the other matrix
coordinates, we label it as $X^{5}$ and use $X^I$ to label all five such
fields.

A more subtle feature of this tachyonic instability is that it is not really
appropriate for us to fix the number of pixels, nor the size of our matrices.
Rather, we must consider the totality of all possible size matrices. Labelling the full reservoir of possible pixels as
$\mathcal{V}_{\text{pixel}}$, we have a direct sum of vector spaces on which our $X^{I}$'s can act:
\begin{equation}
\mathcal{V}_{\text{pixel}}=\underset{N}{%
{\displaystyle\bigoplus}
}\mathcal{V}_{N} \,.
\end{equation}
We can also view the matrices as a power series expanded in non-trivial
functions of the tachyon operator $T = X^5$:
\begin{equation}
  X=\underset{N}{\sum}\Theta_{N}(T)X_{N} \,,
\end{equation}
in the obvious notation. Here, $\Theta_{N}(T)$ is a function of Casimir
invariants built from the tachyons. A natural choice is to view each
$\Theta_{N}(T)$ as a combination of step functions which parameterizes when we
transition from one size matrix to another.

Let us now write down a matrix model which captures the relevant physics of
these pixels. Expanding out the associated DBI\ action and working to leading
order in the various fields, the Lagrangian of the bosonic sector will include
the terms
\begin{equation}
\mathcal{L}_{\text{bosonic}} \supset \text{Tr}\left(  \frac{1}{2} \left\vert D_{t}X^{I}%
\right\vert ^{2}-\frac{\lambda}{4}\left\vert [X^{I},X^{J}]\right\vert
^{2}-V_{\text{tach}}(T)-V_{\text{hop}}(X^{I})+\frac{1}{4!}G_{ijkl}[T, X^{i}]%
X^{j}X^{k}X^{l}\right)  \,, \label{bosonicaction}%
\end{equation}
where the fermionic interaction terms follow from 1D $\mathcal{N}=2$
supersymmetry. In this expression, $D_{t}=\partial_{t}+ i[A_{t},\cdot]$ is a
covariant derivative. Here, we have implicitly specified a target space metric
which includes the tachyon. The commutator potential is the leading order term in the expansion of the
DBI action. We have also included a ``topological term'' associated with
the backreaction of the five-branes as they dissolve into three-form flux. It
is a coupling permitted by all the symmetries of the problem, so we must allow
it. To see how this term comes about, we observe that there is a non-trivial
three-form flux sourced by each pixel. The modified Bianchi identity has the
schematic form
\begin{equation}
  G_{4}=dH_{3}=N\text{ }\delta(x=\text{pixel}) \,,
\end{equation}
where all differential forms have legs in the spatial directions and the delta
function is a four-form indicating the location of our pixels.

The presence of the terms $V_{\text{tach}}(T)$ and $V_{\text{hop}}(X^{I})$
involve interactions which we can at present only model phenomenologically.
The potential $V_{\text{tach}}(T)$ dictates the value of the tachyon operator
as pixel winding/unwinding occurs, and $V_{\text{hop}}(X^{I})$ amounts to a
\textquotedblleft hopping term\textquotedblright\ which relates different
values of the $X^{I}$ as we jump between different size matrices (so that the ambient geometry
probed by the $X^{I}$ is roughly the same at different values of $N$). All we
really require is that this preferentially enforces the condition that
different size matrices canonically embed as the system winds/unwinds.  We
need to enforce the conditions that at a minimum of the potential, $T \cdot T$
is proportional to the identity, and also that the matrices appropriately
embed via $J_{+}X=XJ_{-}$, where a transition to a bigger (resp.  smaller)\
$\mathcal{V}_{N}$ is obtained from raising and lower operators%
\begin{equation}
J_{+}(T):\mathcal{V}_{N}\rightarrow\mathcal{V}_{N+1}\text{ \ \ and \ \ }%
J_{-}(T):\mathcal{V}_{N}\rightarrow\mathcal{V}_{N-1} \,.
\end{equation}
As the notation indicates, we expect that these $J$'s are themselves
non-trivial functions of the tachyon operators in each block, namely the
$T_{N}$'s. A well motivated possibility which is in accord with the analysis
is to take each $J_{+}$ to have entries just along the diagonal.

Even though our proposed action for the pixels has some features which still
need to be developed, we can already see that the matrix model will
preferentially backreact to the geometry of an $S^{3}$, or more precisely a
fuzzy analog of this. Let us note that fuzzy spheres retain the isometries of
the smooth geometry.

To see why we get a fuzzy sphere, consider the physical construction of fuzzy
sphere geometries, e.g. \cite{Castelino:1997rv, Guralnik:2000pb}.  In
reference \cite{Guralnik:2000pb} a rather similar action was put forward to
analyze type IIB\ string theory in the presence of non-BPS\ D0-branes with a
background five-form flux \cite{Guralnik:2000pb}. Next, observe that a
critical point for the classical potential in the $X^{I}$ obeys the
relations%
\begin{equation}\label{foursphere}
\lbrack X^{J},[X^{J},X^{I}]]=\frac{g}{5!}\text{ }\varepsilon_{IJKLM}X^{J}%
X^{K}X^{L}X^{M} \,,
\end{equation}
where $g$ is a constant set by the flux and the strength of the coupling
$\lambda$. 
Singling out the spatial components of the matrix model, we also have%
\begin{equation}\label{threesphere}
\lbrack X^{j},[X^{j},X^{i}]]=\frac{g}{4!}\varepsilon_{ijkl}TX^{j}X^{k}X^{l} \,.
\end{equation}
%
The matrices of equation (\ref{foursphere})
define a fuzzy $S^{4}$ (see reference \cite{Castelino:1997rv}). Moreover, when
$T$ is held fixed with $T \cdot T$ proportional to the identity, the matrices
of equation (\ref{threesphere}) specify a fuzzy $S^{3}$ (see reference
\cite{Guralnik:2000pb}).  The details of this construction are briefly
reviewed in Appendix \ref{app:FUZZ}.  In the case of a fuzzy $S^{4}$ and fuzzy
$S^{3}$, the number of pixels in each case is
\begin{align}
N_{S^{4}}  &  =\frac{(n+3)(n+2)(n+1)}{6},\\
N_{S^{3}}  &  =\frac{(n+3)(n+1)}{4} \,,
\end{align}
where $n$ is a positive integer having to do with how we symmetrize the spinor
representations of $\mathfrak{so}(5)$.

Let us now turn to the physical interpretation.  For restricted values of $N$,
we do indeed produce a fuzzy $S^{3}$, and in the large $N$ limit this goes
over to a smooth spacetime. On the other hand, we also see that there is a
non-trivial relation between the number of pixels and the underlying
symmetrizing factors $n$. This means that not all values of $N$ will produce a
round $S^{3}$ and there will therefore be some inhomogeneities on the $S^{3}$.
This is an extremely small effect. Indeed, in the large $N$ limit, the $S^3$ is smoothed
out so one ought to expect the isometries to re-emerge as $N \rightarrow
\infty$. In this sense the expected anisotropy of the spacetime will be experimentally
negligible. For additional discussion
on potential consequences of physics on a covariant non-commutative
spacetime, see e.g. references \cite{Yang:1947ud, Heckman:2014xha,
VilelaMendes:2019jit}. Note also that working in terms of this $\mathcal{N} = 2$ supersymmetric
quantum mechanics system comes at the price of isometries for our $S^3$. This
addresses a subtle issue which appears in reference \cite{Heckman:2018mxl} in
connection with the absence of systems with two real supercharges on a round $S^3$. At a
generic number of pixels, we do not recover a round fuzzy $S^3$: it has some
``bumps.''

We can also see that the fuzzy sphere configuration has perturbative
instabilities, in line with the classical gravity analysis presented in
\cite{Heckman:2018mxl}. When this occurs, an expansion around the background
values of the fields will involve a kinetic term which depends on the
expansion (or contraction) of the $S^3$. Observe that in the expanding phase,
the strength of the kinetic term starts to become stronger while in the
collapsing phase the potential energy terms start to dominate.

\subsection{De Sitter Thermodynamics}

In the previous sections we presented a cosmological model based on the
winding and unwinding of pixels in our spacetime. In the rare case where $N$
does note change the expanding branch asymptotically approaches de Sitter space. It
should therefore be possible to understand the de Sitter entropy and
temperature in such a background. In principle, this should follow directly
from the matrix quantum mechanics presented earlier, but as there are still
several elements outside our control, we simply make some general qualitative
remarks on the expected $N$ scaling, and how this shows up in the matrix
quantum mechanics system. Note that on the expanding branch, we expect
pixel-pixel interactions will be diluted away.

To set the stage, let us first recall the $N$ scaling for the de Sitter
entropy and temperature. These are given by \cite{Gibbons:1977mu}
\begin{equation}
T_{dS}=\frac{1}{2\pi\ell_{dS}} \quad \text{and} \quad S_{dS}=\frac{A}{4G_{N}%
}=\frac{4\pi\ell_{dS}^{2}}{4G_{N}} \,,
\end{equation}
where the de Sitter radius is related to the cosmological constant via%
\begin{equation}
  \Lambda=\frac{3}{\ell_{dS}^{2}} \,.
\end{equation}
In string units, we therefore observe the following $N$ scaling:
\begin{equation}
  T_{dS}\sim\frac{1}{\sqrt{N}} \quad \text{and} \quad S_{dS}\sim N \,.
\end{equation}

This $N$ scaling matches up with what an observer probing a background
configuration of pixels will experience. Our analysis is quite similar in
spirit to the study of black holes in M(atrix) theory \cite{Banks:1997hz,
Klebanov:1997kv, Banks:1997tn}. First of all, we have precisely $N$
indistinguishable pixels, and we are expanding the matrix model around a
background configuration where we have an $S^{3}$. This means that many of the
off-diagonal modes will not be free to independently fluctuate, so it is more
appropriate to just count the dominant fluctuations around the diagonal. For
this reason we expect all entropic quantities to scale extensively in the
number of pixels, not the number of entries in each matrix (which would have
instead been of order $N^{2}$). Indeed, the number of pixels sets the
\textquotedblleft volume\textquotedblright\ for our system:
\begin{equation}
  S_{\text{obs}}\sim N \,.
\end{equation}

The numerical pre-factor is clearly more delicate. It will depend on the
precise way in which Newton's constant is fixed in the compactification, which
is in turn dependent on the details of the $\mathrm{Spin}(7)$ manifold (in the
backreacted limit)\ or the Calabi--Yau fourfold (prior to backreaction).

Consider next the de Sitter temperature. Clearly, there is no
\textquotedblleft actual temperature\textquotedblright\ in our matrix quantum
mechanics, but because we are dealing with an ensemble of random matrices, an
observer moving in such a background will experience an effective temperature
which translates to a thermalization length as would appear in correlation
functions. This is analogous to what is expected in the eigenstate
thermalization hypothesis (see e.g. \cite{PhysRevA.43.2046, PhysRevE.50.888})
though we hasten to add that our observer cannot really access the full
Hilbert space of the pixels, just the lowest Landau level. This is mainly
because the pixels descend from heavy objects of the compactification.

Now, on a fuzzy space, functions are really $N\times N$ matrices, and
consequently all \textquotedblleft local\textquotedblright\ operators which an
observer could use to perform a measurement will also be built up in the same
way. There are $N^{2}$ independent operators we can specify, and the
correlation length for thermalization tells us we can pack points in a given
4D spacetime volume to length of order $\ell_{\text{th}}^{4}$ before
individual measurements in the spacetime will become correlated. Matching
these two quantities, we learn that $N^{2}$ dictates the thermalization length
via%
\begin{equation}
  \frac{\ell_{\text{th}}^{4}}{\ell_{\ast}^{4}}\sim N^{2} \,,
\end{equation}
or, upon identifying $\ell_{\text{th}}^{-1} \sim T_{\text{obs}}$, we get an
observed temperature of order (in string units)
\begin{equation}
  T_{\text{obs}}\sim\frac{1}{\sqrt{N}} \,,
\end{equation}
so again, we obtain the proper $N$ scaling. As in the case of the entropy
scaling, a more precise match will require additional details of the matrix
quantum mechanics. It is at least encouraging, however, to see simple crude
estimates appear without contrivance.

As a final comment, we also note that one can perform a related computation of
entanglement entropies between the two hemispheres of a fuzzy sphere, and
again, one observes an order $N$ scaling \cite{Dou:2006ni,
Karczmarek:2013jca}.  It would be interesting to directly carry out this
calculation in our physical system.

\section{Conclusions} \label{sec:CONC}

In this paper we have studied cosmological observables for F-theory on a
$\mathrm{Spin}(7)$ background. The scenario begins with a static spacetime
which is unstable against perturbations to an expanding phase. In this model,
the cosmological constant depends inversely on a single parameter $N$ which
can be interpreted as the number of pixels building up the macroscopic
spacetime. These pixels can be created or destroyed, and in so doing produce
radiation. This leads to a mild time-dependence in the cosmological constant
that mimics the structure of slow roll inflation, albeit without the use of a
fundamental scalar. It therefore provides a way to enjoy most of the benefits
of inflation without having to deal with some of the more problematic features
of inflationary model building in string theory, and the conceptual issues
related to tuning the scalar's initial conditions. Additionally, the model
predicts a time-varying cosmological constant at late times that plays the
r\^{o}le of dark energy. Dark energy is well-described by the $w_0$--$w_a$
parametrization with $w_a= -3(1+w_0)\Omega_{m,0}/2$. This is compatible with
current bounds and future data would constrain the time-variation of $N$
further.  The vacuum energy density is protected by an ``$\mathcal{N}=1/2$
supersymmetry,'' thus solving the cosmological constant problem. Finite
excitations are non-supersymmetric, which leads to mass splittings between
Standard Model states and their superpartners. We have also developed the
theory of spacetime pixels undergirding this picture, and have shown how to
recover various qualitative elements of de Sitter thermodynamics such as the
entropy and temperature. In the remainder of this section we discuss some
areas for future investigation.

From an observational point of view, we have seen that the model necessarily
predicts some amount of variation in the equation of state for dark energy.
This should be testable in the next few years and will either constrain the
model, or provide confirmation of the main elements. It would be interesting
to see whether the effects of flux jumps/pixel production of gravitational
radiation can also be observed, perhaps through next generation gravitational
wave experiments. In particular, the pixels (or five-branes in the UV) couple
to gravity and some of the radiation produced during the jumps should be
gravitational.

A feature of our model is that the vacuum energy density observed today is
associated with its value at zero temperature, i.e.  after all phase
transitions have occurred. As noted in \cite{Bellazzini:2015wva,Csaki:2018fls,
Han:2018mtj}, this leads to a modification for the equation of state in
neutron stars, which could potentially be observed in gravitational waves
emitted from binary merger events. It would be interesting to seek out other
astrophysical phenomena sensitive to this feature of our model.

We have also presented evidence that even if the Universe initially falls into
a collapsing branch, it eventually tunnels out to the expanding branch. To
really track this appears difficult to arrange using standard worldsheet
techniques, though the quantum mechanics of our spacetime pixels provides a
promising starting point for dealing with this highly quantum regime.  It
would also provide a firmer starting point for numerous issues in quantum
cosmology.

Finally, we have also presented supporting evidence that the backreaction of
five-branes wrapping metastable five-cycles in an F-theory compactification
builds up a 4D\ spacetime with discretized pixels described by a 1D
supersymmetric quantum mechanics model. We have also seen that at least the
$N$ scaling of the de Sitter entropy and temperature can be read off based on
general considerations. It would of course be interesting to also determine the
precise numerical pre-factors for these quantities.

\newpage

\section*{Acknowledgements}

We thank J. Khoury, L. McAllister, and M. Trodden for helpful discussions. The work of JJH, CL
and GZ is supported by NSF CAREER grant PHY-1756996. The work of LL\ is
supported by DOE Award de-sc0013528y. The work of JS is supported by funds
provided to the Center for Particle Cosmology by the University of
Pennsylvania.

\appendix

\section{The Effective Field Theory of Inflation}
\label{sec:inflation}

In this Appendix we briefly review how one can calculate the spectral index
for the slow-roll regime of the effective field theory of inflation developed
by \cite{Cheung:2007st}. Our starting point is the effective action given in
equation \eqref{eq:action1},
\begin{equation}\label{eq:actionagain}
S=\int\dd^4x\sqrt{-g}\left[\frac{M_{\text{pl}}^2}{2} R +
M_{\text{pl}}^2\dot{H}g^{00}-M_{\text{pl}}^2(3H^2+\dot{H})+\cdots\right] \,.
\end{equation}
Now one can always restore the broken time-diffeomorphisms by introducing
suitable St\"{u}ckelberg fields.  In order to accomplish this, we replace $t$
by $t+\pi(x)$ and assign $\pi(x)$ the transformation law under
time-diffeomorphisms ($t\rightarrow t+\xi^0(x)$),
\begin{equation}
  \pi(x)\rightarrow\pi(x)-\xi^0(x) \,,
\end{equation}
so that every term is now diffeomorphism-invariant. The new field $\pi$ is the
Goldstone boson of broke time-diffeomorphisms, and its fluctuations encode the
spectrum of scalar perturbations.  Performing the St\"{u}ckelberg trick on the
action \eqref{eq:actionagain} one finds
\begin{align}\label{eq:action2}
S&=\int\dd^4x\sqrt{-g}\left[\frac{M_{\text{pl}}^2}{2}R -M_{\text{pl}}^2(3H^2(t+\pi)+\dot{H}(t+\pi)\right.\\
&\left.+M_{\text{pl}}^2\dot{H}(t+\pi)\left((1+\dot{\pi})^2g^{00}+2(1+\dot{\pi})g^{0i}+g^{ij}\partial_i\pi\partial_j\pi\right)\right]
\,,
\end{align}
which is diffeomorphism-invariant. By inspection of the kinetic terms for
$\pi$, we see that the canonically normalized field is $\pi_c\sim
M_{\text{pl}}\sqrt{\dot{H}}\pi$.  Now we can relate $\pi$ to the curvature
perturbation $\zeta$ defined via
\begin{equation}
  \dd s^2=-\dd t^2 +a^2(t)\left[(1+2\zeta)\delta_{ij}\right]\dd x^i\dd x^j \,,
\end{equation}
in the unitary gauge ($\pi=0$). Performing a gauge transformation to set
$\pi=0$ (in the action \eqref{eq:action2}) requires us to set $t\rightarrow
t-\pi$.  Since the metric is unperturbed in the $\pi=0$ gauge (we are ignoring
  the mixing with gravity in the spirit of effective field theory), one can
  perform this transformation to find
\begin{equation}
  g_{ij}\rightarrow a^2(t-\pi)\delta_{ij}  = a^2(t)(1-2H\pi)\delta_{ij} \,,
\end{equation}
so that $\zeta=-H\pi$. Our theory is now just single-field slow-roll
inflation, for which we know
\begin{equation}
\langle\pi_c(\vec{k}_1)\pi_c(\vec{k}_2)\rangle=(2\pi)^3\delta^{(3)}(\vec{k}_1+\vec{k}_2)\frac{H_\ast^2}{2k_1^3}
\,,
\end{equation}
where subscript asterisks denote quantities evaluated at horizon crossing
($k=aH$). Relating this to $\zeta$ gives us
\begin{equation}\label{eq:PSzeta}
\langle\zeta(\vec{k}_1)\zeta(\vec{k}_2)\rangle=\frac{(2\pi)^3}{k_1^3}\delta^{(3)}(\vec{k}_1+\vec{k}_2)\frac{H_\ast^2}{4\ve_\ast
M_{\text{pl}}^2} \,.
\end{equation}
The tilt of the power spectrum is given by standard computations (see
\cite{Cheung:2007st} and references therein):
\begin{equation}
  n_s-1=-2\ve-\eta \,.
\end{equation}
Another important quantity is the amplitude $A_s$ of the power spectrum, as
well as the spectral index $n_s$ as implicitly defined via
\begin{equation}\label{eq:PSzeta2}
\langle\zeta(\vec{k}_1)\zeta(\vec{k}_2)\rangle = \frac{2 \pi^2}{k_1^3} \delta^{(3)}(\vec{k}_1+\vec{k}_2)
\times  A_s \left(\frac{k_1}{k_\ast}\right)^{n_s-1} \,.
\end{equation}

\section{Discontinuous Radiation and Dark Energy Jumps} \label{app:DISCO}

In this Appendix we  show how energy densities behave under discontinuous
jumps in the number of flux units. We will take the jump to happen at some
time $t = t_\ast$ and take the Universe before the jump to have only
cosmological constant contributing to the energy density. After the jump some
energy density will have the form of radiation and some remnant cosmological
constant. Therefore we can parametrize the behavior of the energy density
close to the jump as
\begin{align}
\label{eq:rhojump}\rho = \rho_{\text{rad}} + \rho_\Lambda = \frac{\gamma \theta (t-t_\ast)}{a^4} + \frac{1}{8 \pi G_N} \left[\Lambda_0 - \theta(t-t_\ast) \delta \Lambda\right]\,.
\end{align}
Here $\theta(t)$ is the Heaviside step function that we used to model the
instantaneous jump. While in general we express $\delta \Lambda$ in terms of
the change of flux units we can keep it more general in this analysis. In
writing \eqref{eq:rhojump} we were not able to write directly a relation
between the radiation energy density after the jump and the change in
cosmological constant; we will be able to fix the relation between the two
quantities using the Friedmann equations. We assume that close to the jump we can
safely treat the system's dynamics as well approximated by its classical
dynamics. Our main tool will be the conservation of the stress-energy tensor
$\nabla_\nu T^{\mu \nu} =0$ which in a FLRW background takes the form
\begin{align}
\dot \rho = -3 H (\rho + P)\,.
\end{align}
Using the relation $P_{\text{rad}} = 1/3 \rho_{\text{rad}} $ and $P_\Lambda =
- \rho_\Lambda$ we get that using the parametrization \eqref{eq:rhojump}
conservation of the stress-energy tensor becomes
\begin{align}
\label{eq:cont}\frac{\gamma}{a_\ast^4} = \frac{1}{8 \pi G_N} \delta \Lambda\,,
\end{align}
where $a_\ast = a(t_\ast)$. Using this relation we may try to assess the
behavior of the scale factor and its derivatives across the jump as
constrained by Friedmann equations. Using the Friedmann equation
\eqref{eq:Friedm1} one can easily show that the first derivative of the scale
factor is continuous across the jump. The remaining Friedmann equation is
automatically satisfied because of the conservation of the stress-energy
tensor, however in this case it is possible to see that the second derivative
cannot be continuous across the jump. To be more concrete we can give an
analytic solution of Friedmann equations in the case where $\delta \Lambda =
\Lambda_0$ meaning that after the jump radiation is the only component of the
energy density. In this case the solution is rather simple and takes the form
\begin{align}
a(t) = \left\{\begin{array}{l l}e^{\sqrt{\frac{\Lambda_0}{3}} (t-t_\ast)}\left( \frac{8 \pi G_N \gamma}{\Lambda_0}\right)^{\frac{1}{4}} & t\leq t_\ast\\[3mm]
\left(8 \pi G_N\gamma\right)^{\frac{1}{4}}\sqrt{\frac{2}{\sqrt{3}} (t-t_\ast) +\frac{1}{\sqrt{\Lambda_0}}} & t\geq t_\ast\end{array}\right.\,.
\end{align}
Here $\gamma$ is fixed by the relation \eqref{eq:cont}. With this solution one
can see that the scale factor and its first derivative are continuous at
$t=t_\ast$ but the second derivative has a discontinuity:
\begin{align}
\ddot a (t_\ast^-)-\ddot a (t_\ast^+) = \frac{2}{3} \left(8 \pi G_N \Lambda_0 \gamma\right)^{\frac{1}{4}}\,.
\end{align}

\section{Geometry of Fuzzy Spheres} \label{app:FUZZ}

In this Appendix we briefly we review the geometry of the fuzzy spheres
$S^{4}$ and $S^{3}$. We refer the interested reader to \cite{Castelino:1997rv,
Guralnik:2000pb} for additional discussion. An important feature of such
geometries is that they retain all the isometries of their classical
counterparts. This follows from the adjoint action on the matrix coordinates
used to build the sphere.

Consider first the case of an $S^{4}$. The main idea here is that we make an
ansatz in which the $X^{I}$'s are proportional to symmetrizations of the gamma
matrices of $\mathfrak{so}(5)$. Denoting the spinor representation by
$\mathcal{S}$, we get the fuzzy four-sphere by forming $n$-fold symmetrized
products $\text{Sym}^{n}(\mathcal{S})$ and introducing matrix coordinates%
\begin{equation}
G^{I}=\left(  \Gamma^{I}\otimes1\otimes...\otimes1+1\otimes...\otimes
1\otimes\Gamma^{I}\right)  _{\text{symm}} \,,
\end{equation}
and we set%
\begin{equation}
  X^{I}=\alpha G^{I} \,,
\end{equation}
with $\alpha$ a constant of proportionality dictated by the physical
potential/size of the $S^{4}$. The size of the matrices is in this case
controlled by the representation theory of $\mathfrak{so}(5)$ to be%
\begin{equation}
  N_{S^{4}}=\frac{(n+3)(n+2)(n+1)}{6} \,.
\end{equation}
Observe that the commutators of two $G$'s build a generator of $\mathfrak{so}(5)$
rotations, namely%
\begin{equation}
M^{IJ}=\frac{1}{2}[G^{I},G^{J}]\text{ \ \ so that \ \ }\left[  M^{IJ}%
,G^{K}\right]  =2(\delta^{JK}G^{I}-\delta^{IK}G^{J}) \,,
\end{equation}
so one can check that this ansatz does indeed solve equation
(\ref{foursphere}). Additionally, the $G$'s satisfy the relation%
\begin{equation}
  G^{I}G^{I}=n(n+4) \,,
\end{equation}
namely we really do get the geometry of an $S^{4}$.

To get a fuzzy $S^{3}$ we restrict to the equator of the $S^{4}$, that is, we
single out one direction and identify it with $\Gamma^{5}$ in the 4D Dirac
algebra. Clearly, this is to be identified with the tachyonic mode of the
matrix model introduced in section \ref{sec:PIXEL}. In this
case, the spinor representation of $\mathfrak{so}(5)$ splits into a positive
and negative chirality spinor of $\mathfrak{so}(4) =
\mathfrak{su}(2)_{+}\times \mathfrak{su}(2)_{-}$:
\begin{align}
\mathfrak{so}(5)  &  \rightarrow \mathfrak{su}(2)_{+}\times \mathfrak{su}(2)_{-}\\
\mathcal{S}  &  \rightarrow\mathcal{S}_{+}\oplus\mathcal{S}_{-} \,,
\end{align}
and we must project onto a reducible representation $\mathcal{R}$ of $\mathfrak{so}(4)$
using a linear operator $P_{\mathcal{R}}$ so that the projected matrices%
\begin{equation}
  \widehat{G}^{i}=P_{\mathcal{R}}G^{i}P_{\mathcal{R}} \,,
\end{equation}
define a fuzzy $S^{3}$, namely we require $\widehat{G}^{i}\widehat{G}^{i}$ is
proportional to the identity. As explained in \cite{Guralnik:2000pb} this can
be arranged for the reducible representation\footnote{We denote a
  two-dimensional representation as having spin $1/2$.}%
\begin{equation}
\mathcal{R}=\left(  \frac{n+1}{4},\frac{n-1}{4}\right)  \oplus\left(
\frac{n-1}{4},\frac{n+1}{4}\right)  .
\end{equation}
So in this case, the number of pixels on the fuzzy $S^{3}$ is%
\begin{equation}
  N_{S^{3}}=\frac{(n+3)(n+1)}{4} \,.
\end{equation}

\newpage

\bibliographystyle{utphys}
\bibliography{FInflation}

\end{document}